\documentclass[12pt]{article}
\usepackage{arxiv}
\usepackage{url} 
\usepackage{amsmath,amsfonts,amssymb}
\usepackage{bm}
\usepackage{epsf,graphicx} 
\usepackage{lineno}
\usepackage{apacite}
\usepackage{natbib}

\mathchardef\mhyphen="2D

\begin{document}

%
%


\title{A theoretical framework of chorus wave excitation}

%
%





\author{
  F. Zonca\thanks{\texttt{fulvio.zonca@enea.it}}\\
  Center for Nonlinear Plasma Science and C.R. ENEA Frascati - C.P. 65, 00044 Frascati, Italy
  \And
  X. Tao\\
  Department of Geophysics and Planetary Sciences, University of Science and Technology of China, Hefei, China\\
  \And
  L. Chen\\
  Institute of Fusion Theory and Simulation, Department of Physics, Zhejiang University, Hangzhou, China
}










\maketitle

\newcommand{\cb}{\color{blue} }
\newcommand{\cg}{\color{green} }
\newcommand{\cm}{\color{magenta} }
\newcommand{\crr}{\color{red} }
\newcommand{\laeq}{\raisebox{-.7ex}{$\stackrel{\textstyle<}{\sim}$}\ }
\newcommand{\gaeq}{\raisebox{-.7ex}{$\stackrel{\textstyle>}{\sim}$}\ }
\newcommand{\exgra}{{\rmfamily\itshape\mdseries e.g.}}
\newcommand{\idest}{{\rmfamily\itshape\mdseries i.e.}}
\newcommand{\etal}{{\rmfamily\itshape\mdseries et al.}}

%
%


\begin{abstract}
We propose a self-consistent theoretical framework of chorus wave excitation, which
describes the evolution of the whistler fluctuation spectrum as well as the supra-thermal 
electron distribution function. The renormalized hot electron response is cast in 
the form of a Dyson-like equation, which then leads to evolution 
equations for nonlinear fluctuation growth and frequency shift. This approach allows us
to analytically derive for the first time exactly the same expression for the chorus chirping rate originally proposed
by \cite{Vomvoridis1982}. Chorus chirping is shown to correspond to 
maximization of wave particle power
exchange, where each individual wave belonging to the whistler wave packet 
is characterized by small nonlinear frequency shift. 
We also show that different interpretations of chorus chirping
proposed in published literature have a consistent reconciliation within the present theoretical 
framework, which further illuminates the analogy with similar phenomena in fusion plasmas
and free electron laser physics.

\end{abstract}

%
%

%


%
%
%
%

\section{Introduction}
Chorus waves are whistler mode waves with frequency ($\omega$) 
typically between one tenth of and the electron cyclotron frequency ($\Omega$) \citep[see, e.g.,][]{Tsurutani1974,Burtis1976}. These waves have been demonstrated 
to play important roles in energetic electron dynamics in the terrestrial magnetosphere. Chorus waves are responsible for the acceleration 
of a few hundred keV electrons to the MeV energy range, leading to the enhancement of MeV electron fluxes in the outer radiation belt 
during geomagnetically disturbed times \citep{Horne1998, Horne2005b,Chen2007,Bortnik2007c,Reeves2013,Thorne2013b}. Furthermore,  
scattering of a few hundred eV to a few keV electrons by chorus waves into the atmosphere has been shown to be the dominant process in 
the formation of energetic electron pancake distributions \citep{Tao2011b}, diffuse aurora \citep{Thorne2010}, and pulsating 
aurora \citep{Miyoshi2010,Nishimura2010}. 

Chorus waves consist of quasi-coherent discrete elements with frequency chirping. 
In the terrestrial magnetosphere, the frequency of a chorus 
element may vary by a few hundred Hz to a few kHz in less than a second. 
Previous studies have established that coherent nonlinear wave particle 
interactions play a key role in the frequency chirping, and have demonstrated that the 
chirping rate for rising tone chorus is proportional to the 
wave amplitude. Using a series of simulations, \citet{Vomvoridis1982} argued that, 
to maximize wave power transfer, the frequency 
chirping rate and the wave amplitude for parallel propagating chorus waves is related by
\begin{align}
  \label{eq:1}
  \frac{\partial \omega}{\partial t} =  R \left(1-\frac{v_r}{v_g}\right)^{-2} \omega_{tr}^2,
\end{align}
with $R=1/2$. Here $v_r$ is the cyclotron resonant velocity, $v_g$ is the wave group velocity, 
and $\omega_{tr}^2 = k v_\perp e \delta B/(mc)$ 
with $v_\perp$ the perpendicular velocity, $k$ the wave number, and $\delta B$ the wave amplitude. 
Theoretical interpretation of Eq. (\ref{eq:1}) was proposed by \citet{Trakhtengerts2004} and \citet{Demekhov2011},
based on the assumption that a chorus element is formed as a succession of sidebands separated from each other by 
the trapping frequency $\omega_{tr}$ over timescales $2\pi/\omega_{tr}$. 
Meanwhile, for optimum cyclotron power exchange of electrons with a whistler wave in an inhomogeneous magnetic field,
the growth rate is determined by a Backward Wave Oscillator (BWO) condition. Good agreement from comparisons of these chorus sweeping rate predictions with
observations was reported by \citep{Trakhtengerts2004,Macusova2010,Tao2012b}. 
Another interpretation of Eq. (\ref{eq:1}) was proposed by \citet{Omura2008}, by
assuming a constant value for the 
phase space density of phase-trapped electrons, and demonstrating that the resonant 
current density in the direction of wave 
electric field maximizes at $R=0.4$, consistent with Eq. (\ref{eq:1}) of \citet{Vomvoridis1982}. 
Equation (\ref{eq:1}) has been verified 
by several different particle-in-cell (PIC) type simulations \citep{Katoh2011a,Katoh2013,Hikishima2012,Tao2017a,Tao2017b} and an 
observational study \citep{Cully2011}.  More recently, \citet{Mourenas2015} suggested using the nonlinear chorus growth rate from \citet{Shklyar2009}, based on contributions from both trapped and untrapped resonant particles, to derive an analytical estimate of the value of $R = S^*$ maximizing this nonlinear growth rate. They obtained $R=S^* = 3/5$ in the case of oblique chorus waves.

Despite of Eq. (\ref{eq:1}) being a huge success, its derivation was based either on  
simulations \citep{Vomvoridis1982} or 
by assuming either a given behavior of the fluctuation spectrum \citep{Trakhtengerts2004,Demekhov2011}, or 
a specific form of the distribution function for phase trapped \citep{Omura2008} and/or untrapped \citep{Shklyar2009,Mourenas2015} particles. 
Besides , Eq. (1) was derived assuming the pre-existence of frequency chirping
 \citep{Vomvoridis1982}. The reason for frequency chirping of chorus was explained by 
 \cite{Omura2011} due to the nonlinear current parallel to the wave magnetic field ($J_B$), 
 which causes frequency shift.
In this paper, we propose a new first principle based theoretical 
framework for chorus wave excitation that addresses the dynamic evolution of the fluctuation spectrum
and its interaction with trapped as well as untrapped resonant particles on the same footing. This {\em self-consistent} analysis is a novelty of the
present approach with respect to previous studies. By keeping the dominant long-term 
nonlinear response in the distribution function, 
we obtain an equation for the evolution of the distribution function of hot electrons in the form of a 
Dyson-like equation \citep{Dyson1949, Schwinger1951}. 
This model \citep{Chen2016, Zonca2015} explains chirping as a result of the dynamic nonlinear spectrum evolution 
due to coherent excitation of a narrow fluctuation spectrum that is shifting in time out of a broad and dense whistler wave spectrum. 
Furthermore, it demonstrates the ballistic propagation of resonant structures in the hot electron phase space
\citep{Chen2016, Zonca2015}, and analytically shows that 
maximization of wave power transfer leads to $R=1/2$ 
and Eq. (\ref{eq:1}), fully coincident with previous results of \citet{Vomvoridis1982}. 
At last, the present theoretical framework illuminates why 
the original approach by \cite{Omura2011}, based on the nonlinear frequency shift due to
$J_B$, yields the correct estimate of chorus chirping rate starting from a different 
perspective.

In the present analysis, we focus on the nonlinear dynamics of 
phase space structures of correlated electrons, which are due to nonlinear wave particle interactions that 
predominantly occur ``in the downstream of equator'', after the whistler wave packets have traveled through
the supra-thermal electron source region localized near the equator itself. In this respect, we construct the
theoretical framework that underlies the numerical simulation analysis of \citet{Tao2017a}, where we showed
that the time scale of chorus nonlinear dynamics is $\sim {\cal O}(2\pi/\omega_{tr})$, characteristic of non-perturbative
wave-particle interactions \citep{Chen2016, Zonca2015}, and that the wave growth is mainly due to phase bunched electrons. This work
is a more in depth analysis based on earlier preliminary theoretical approach \citep{Zonca2017}. 
The present analysis can also be considered as theoretical building block for a recent simulation work
\citep{Tao2021}, which proposes a novel phenomenological interpretation for chorus, called the ``Trap-Release-Amplify'' (TaRA) model.
The TaRA model establishes a connection between the upstream and downstream of equator regions
in chorus dynamics, and shows that 
phase-locked electrons in the upstream region selectively amplify wave packets with a chirping rate that is fully consistent
with the Helliwell analysis for a nonuniform background magnetic field \citep{Helliwell1967}. Meanwhile, in the 
downstream region, the nonlinear wave-particle analysis in the TaRA model yields the chorus chirping 
expression of Eq. (\ref{eq:1}), consistent with \citet{Vomvoridis1982} as well as with former \citep{Tao2017a,Zonca2017}
and present analyses.

The structure of the paper is as follows. Section \ref{sec:theory} discusses the present novel theoretical framework
based on the self-consistent solution of wave equations (Sec. \ref{sec:waveequations}) and of 
nonlinear phase-space dynamics (Sec. \ref{sec:phasespace}). Reduced model equations for the 
nonlinear evolution of spectral intensity and phase shift are presented in Sec. \ref{sec:reddyson}.
These are then applied to the investigation of chorus excitation and nonlinear dynamics in 
Sec.  \ref{sec:choruschirp}. Finally, Sec. \ref{sec:summary} is devoted to discussion and 
concluding remarks. Four appendixes are further devoted to detailed derivations for
interested readers.

\section{Theory}
\label{sec:theory}

Let us adopt the standard hybrid approach, where the Earth's magnetosphere plasma is assumed 
to consist of a neutralizing cold thermal ion background and
a ``core'' (c) and ``hot'' (h) electron components. From Amp\`ere's law, and separating the current 
density perturbation in $\delta \bm J$ in ``core'' (c) and ``hot'' (h) components, for parallel
propagating transverse electromagnetic waves
we have 
\begin{equation}
\left( 1 - \frac{k^2 c^2}{\omega^2} \right) \delta \bm E_\perp  + \frac{4\pi i}{\omega} \delta \bm J_c \equiv \bm \epsilon_\perp \cdot \delta \bm E_\perp - \frac{k^2 c^2}{\omega^2} \delta \bm E_\perp = - \frac{4\pi i}{\omega} \delta \bm J_h \; , \label{eq:amperemod0}
\end{equation}
where, considering ``core'' electrons as a cold fluid with density $n$, $\bm \epsilon_\perp$ is the 
usual cold plasma dielectric tensor 
\begin{equation}
\bm \epsilon_\perp \cdot \delta \bm E_\perp = \left( 1 + \frac{\omega_p^2}{\Omega^2 - \omega^2}\right) \delta \bm E_\perp
- i \frac{\Omega}{\omega} \frac{\omega_p^2}{\Omega^2 - \omega^2} \hat {\bm z} \times \delta \bm E_\perp \; , \label{eq:coldplasma}
\end{equation}
with $\hat {\bm z}$ the unit vector along the Earth's magnetic field and, adopting standard notation, $\omega_p^2 = 4\pi ne^2/m$ is the electron plasma frequency 
and $\Omega = e B/(mc)$ is the electron cyclotron frequency, with $e$ the positive electron 
charge and $m$ the electron mass.

For a typical rising tone chorus event that we are addressing here, the characteristic nonlinear time and 
duration are much shorter than the time it takes for a whistler wave to propagate from the equator to 
either southern or northern ionospheres. Thus, we assume a wave packet description for chorus,
which has a dense (nearly continuous) spectrum and is nearly degenerate with a parallel propagating whistler wave
with right circular polarization; 
i.e., $\bm k = \hat{\bm z} k$, $\delta E_z = \delta B_z =0$, $\delta E_y = i \delta E_x$ and
$\delta B_y = i \delta B_x$, with $\delta B_y = (ck/\omega) \delta E_x$. This yields:
\begin{equation}
\bm \epsilon_\perp \cdot \delta \bm E_\perp \simeq \left( 1 + \frac{\omega_p^2}{\omega(\Omega - \omega)}\right) \delta \bm E_\perp \; . \label{eq:coldchorus}
\end{equation}
Thus, the problem of transverse chorus wave packet interacting with hot electrons can be approximately cast as \citep{Nunn1974,Omura1982,Omura2008,Omura2011}
\begin{equation}
\left( 1 + \frac{\omega_p^2}{\omega(\Omega - \omega)}\right) \delta \bm E_\perp  - \frac{k^2 c^2}{\omega^2} \delta \bm E_\perp = - \frac{4\pi i}{\omega} \delta \bm J_h 
\; , \label{eq:master} \end{equation}
where the right hand side can be formally treated as a perturbation to the lowest order propagation of the 
whistler wave packet due to the low density of hot electrons.

\subsection{Wave Equations}
\label{sec:waveequations}

Let us introduce the whistler wave dielectric constant, $\epsilon_w$, and dispersion function, 
$D_w$, such as
\begin{equation}
\epsilon_w = 1 + \frac{\omega_p^2}{\omega(\Omega - \omega)} \; , \;\;\;\;\; D_w = \epsilon_w  - \frac{k^2 c^2}{\omega^2} \; .  \label{eq:chorusdef}
\end{equation}
The elements of the whistler wave packet can be written as
\begin{equation}
\delta \bm E_\perp (z,t)  = \frac{1}{2} \sum_k \left(  e^{i S_k(z,t)} \delta \bar{\bm E}_{\perp k}(z,t) + c.c. \right) \; , 
\label{eq:choruswp}
\end{equation}
with $c.c.$ denoting the complex conjugate, denoted in the following with a $^*$ superscript, and the eikonal $S_k$ is defined such that $\omega_k = - \partial_t S_k$, $k = \partial_z S_k$, which
satisfy the WKB dispersion relation
\begin{equation}
D_w(z,k(z),\omega_k) = 0 \; . \label{eq:chorusdisp}
\end{equation}
Meanwhile, letting 
\begin{equation}
\delta \bar{\bm E}_{\perp k}(z,t) = \hat{\bm e} |\delta \bar{\bm E}_{\perp k}(z,t)| e^{i \varphi_k (z,t)} \; , \label{eq:ampphase}
\end{equation}
with $\hat{\bm e}$ the polarization vector defined such $\hat{\bm e} \cdot \hat {\bm z} = 0$ and $\hat{\bm e} \cdot \hat{\bm e}^* = 1$; 
and introducing wave intensity or action as
\begin{equation}
I_k(z,t) \equiv \left|\partial D_w/\partial k\right| |\delta \bar{\bm E}_{\perp k}(z,t)|^2 \; , \label{eq:waveaction}
\end{equation}
the evolution equation for $I_k(z,t)$ is \citep{Bernstein1977,McDonald1988}
\begin{equation}
\left( \frac{\partial}{\partial t} + v_{gk} \frac{\partial}{\partial z} \right) I_k(z,t) = 2 \gamma_k I_k(z,t) \; , \label{eq:actionevolve}
\end{equation}
where $v_{gk} = - (\partial D_w/\partial \omega_k)^{-1}\partial D_w/\partial k$ is the wave packet group velocity and
\begin{equation}
\gamma_k = - \frac{D_{Ak}^1}{\partial D_w/\partial \omega_k}
\end{equation}
represents the wave packet driving rate due to the hot electrons. 
In fact, noting $\partial_z\partial_k D_w = 0$ \citep{McDonald1988},
\begin{equation}
D_{A k}^1 = \mathbb I{\rm m} \left( \frac{4\pi i}{\omega_k} \frac{ \delta \bar{\bm J}_{h k} \cdot \delta \bar{\bm E}_{\perp k}^*}{|\delta \bar{\bm E}_{\perp k}(z,t)|^2} \right)\; .
\label{eq:antiHD}
\end{equation}
Meanwhile, the phase shift $\varphi_k (z,t)$ is given by \citep{Bernstein1977,McDonald1988}
\begin{equation}
 \left( \frac{\partial}{\partial t} + v_{gk} \frac{\partial}{\partial z} \right) \varphi_k (z,t) = \frac{D_{R k}^1}{\partial D_w/\partial \omega_k} \; ,  \label{eq:phaseevolve}
\end{equation}
where 
\begin{equation}
D_{R k}^1 = \mathbb R{\rm e} \left( \frac{4\pi i}{\omega_k} \frac{ \delta \bar{\bm J}_{h k} \cdot \delta \bar{\bm E}_{\perp k}^*}{|\delta \bar{\bm E}_{\perp k}(z,t)|^2} \right) \; . 
\label{eq:realHD}
\end{equation}
Thus, all relevant nonlinear physics is included in the wave particle interaction and the 
two functions $W(z, t, \omega)$ and $\Gamma(z, t, \omega)$, representing, respectively, the phase shift and driving rate due to 
hot electrons;
\begin{equation}
W(z, t, \omega)+i \Gamma(z, t, \omega) \equiv-\left.\frac{4 \pi i}{\omega \partial D_{w} / \partial \omega} \frac{ \delta \bar{\bm J}_{h k} \cdot \delta \bar{\bm E}_{\perp k}^*}{|\delta \bar{\bm E}_{\perp k}(z,t)|^2}\right|_{k=K(z, \omega)}  \; , \label{eq:WGamma}
\end{equation}
where $k=K(z,\omega)$ is obtained from the solution of the linear dispersion relation,  $D_w=0$.
Noting that, in the complex wave representation adopted here, $\delta \bar{\bm J}_{h k} \cdot \delta 
\bar{\bm E}_{\perp k}^* = - i (\omega/kc) \delta \bar{\bm J}_{h k} \cdot \delta \bar{\bm B}_{\perp k}^*$ 
for the considered right circularly polarized parallel propagating whistler wave packet, Eqs. (\ref{eq:waveaction}) -- (\ref{eq:WGamma}) 
coincide with those adopted by \cite{Nunn1974} and \cite{Omura2011}. With the definition of  $W(z, t, \omega)$ and $\Gamma(z, t, \omega)$ as in Eq. (\ref{eq:WGamma}),
the right hand side of Eqs. (\ref{eq:actionevolve}) and (\ref{eq:phaseevolve}) become, respectively, $2 \Gamma(z, t, \omega) I (z, t, \omega)$
and $- W(z, t, \omega)$, where $I (z, t, \omega) \equiv I_k(z,t)|_{k=K(z,\omega)}$ and the group velocity
on the left hand side has to be interpreted as $v_{g \omega}(z) = v_{g k}(z)|_{k=K(z,\omega)}$.

Equation (\ref{eq:WGamma}) can be rewritten expressing the wave particle power transfer in terms of the hot electron response. In fact,
noting the hot electron right hand cyclotron motion in the ambient magnetic field, 
\begin{equation}
\bm v  \cdot \delta \bar{\bm E}_{\perp k}^* = - i v_\perp e^{-i\alpha} \delta \bar E_k^*\; , \label{eq:perpex}
\end{equation}
where $v_\perp$ (and $v_\parallel$) indicate
the perpendicular (and parallel, respectively) velocity with respect to the ambient 
Earth's magnetic field, $\alpha$ is the gyrophase ($\dot \alpha = \Omega$); and 
we have denoted $\delta \bar E_k \equiv \left(\delta \bar{\bm E}_{\perp k}\right)_x$ for brevity. 
Meanwhile, hot electron response can be represented as
\begin{equation}
f (\bm v, z,t)  =  f_0 (\bm v, z, t) + \frac{1}{2} \sum_k \left(  e^{i S_k(z,t) + i \alpha} \delta \bar f_k (\bm v, z,t) + c.c. \right) \; , 
\label{eq:elef}
\end{equation}
while the hot electron perpendicular current is given by
\begin{equation}
\delta \bm J_h (z,t)  = \frac{1}{2} \sum_k \left(  e^{i S_k(z,t)} \delta \bar{\bm J}_{h k}(z,t) + c.c. \right) \; . 
\label{eq:Jhwp}
\end{equation}
Thus, combining Eqs. (\ref{eq:perpex}) to (\ref{eq:Jhwp}), we have
\begin{eqnarray}
- \left.\frac{4 \pi i}{\omega} \frac{ \delta \bar{\bm J}_{h k} \cdot \delta \bar{\bm E}_{\perp k}^*}{|\delta \bar{\bm E}_{\perp k}(z,t)|^2}\right|_{k=K(z, \omega)}
& = & \left\langle \frac{4\pi e}{\omega} \frac{v_\perp }{2} \frac{\delta \bar E_k^* \delta \bar f_k}{|\delta \bar E_k|^2} \right\rangle
\nonumber \\
& = & \frac{\omega_p^2}{n \omega} \left\langle \frac{v_\perp^2/2}{\Omega + 
k v_\parallel - \omega}  
\left[ \frac{k}{\omega} \frac{\partial f_0}{\partial v_{ \|}}+\left(1-\frac{k v_{ \|}}{\omega}\right) \frac{1}{v_{\perp}} \frac{\partial f_0}{\partial v_{\perp}} \right] \right\rangle \; . \label{eq:powerexchange}
\end{eqnarray}
Here, we have noted that $|\delta \bar{\bm E}_{\perp k}(z,t)|^2 = 2 |\delta \bar E_k|^2$,
angular brackets $\left\langle ... \right\rangle$ denote
velocity space integration, and assumed Eq. (\ref{eq:dbarfkvla0}) for expressing $\delta \bar f_k$, 
which will be derived in Sec. \ref{sec:phasespace}. Furthermore, it is important to emphasize that 
$f_0$ is the $k=0$ component of the hot electron distribution function, which evolves in time due to the 
nonlinear wave particle interactions. Thus, $f_0$ is the ``equilibrium'' distribution function assumed initially only at $t=0$. 

\begin{figure}[t]
\centerline{\resizebox{\textwidth}{!}{\includegraphics{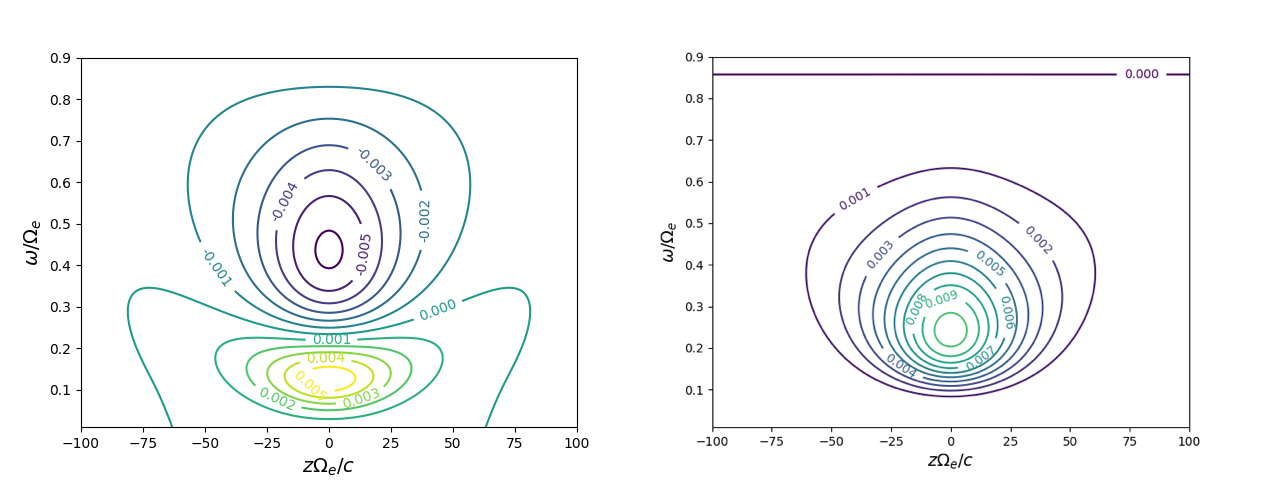}}} \
\vspace*{-2em}\newline\noindent(a) \hspace*{0.48\linewidth} (b)
\caption{Contour plots of $W(z, t=0, \omega)/\Omega_e$ (a) and $\Gamma(z, t=0, \omega)/\Omega_e$ (b) are shown
for normalized parameters $\omega_{p}/\Omega_e=5$, $n_e/n = 6 \times 10^{-3}$, $w_{\parallel e} = 0.2 c$, $w_{\perp e} = 0.53 c$,
$\xi = 8.62 \times 10^{-5} \Omega_e^2/c^2$.}
\label{fig:WGamma}
\end{figure}

Since the source region of chorus is localized near the equator, we follow the usual practice of assuming a model of Earth's dipole magnetic field
in the form $B=B_e (1 + \xi z^2)$ \citep{Helliwell1967}, with $B_e$ representing the magnetic field strength at the equator and 
$\xi^{-1/2}$ the non-uniformity scale length. Thus, we take a model $\Omega = \Omega_e (1 + \xi z^2)$
with $\Omega_e$ the (non-relativistic) electron cyclotron frequency at the equator. Meanwhile, following \cite{Tao2014b},
we assume an ``initial'' hot electron $f_0$ in the form of a bi-Maxwellian
\begin{equation}
\left. f_0\right|_{t=0} = \frac{n_0}{(2\pi)^{3/2} w_{\parallel} w_{\perp}^2} \exp \left( - v_\perp^2/(2w_{\perp}^2) - v_\parallel^2/(2w_{\parallel}^2)  \right) \label{eq:f0eq}
\; ,
\end{equation}
where $n_0 = \zeta^2 n_e$, $w_{\parallel} = w_{\parallel e}$, $w_{\perp} = \zeta w_{\perp e}$, and 
$\zeta^{-2} = 1 + A \xi z^2/(1+ \xi z^2)$, with $A \equiv w_{\perp e}^2/w_{\parallel e}^2 - 1>0$ the anisotropy index computed at the equator (cf. \ref{app:linear}).
Contour plots of the ``initial'' (linear) functions $W(z, t=0, \omega)$ and $\Gamma(z, t=0, \omega)$ are given in
Fig. \ref{fig:WGamma}, for normalized parameters $\omega_{p}/\Omega_e=5$, $n_e/n = 6 \times 10^{-3}$, $w_{\parallel e} = 0.2 c$, $w_{\perp e} = 0.53 c$,
$\xi = 8.62 \times 10^{-5} \Omega_e^2/c^2$ \citep{Tao2014b}.
From Eqs. (\ref{eq:WGamma}) and (\ref{eq:powerexchange}), it can be shown that both $W(z, t, \omega)$ 
and $\Gamma(z, t, \omega)$ scale as $\sim \zeta^4$,
which accounts for most of the spatial non-uniformity of hot electron response, characterized by the length 
scale $(A \xi)^{-1/2}$ as clearly illustrated by Fig. \ref{fig:WGamma} (cf. \ref{app:linear}). Thus, at the leading order,
hot electrons can be considered as a non-uniform source
neglecting magnetic field non-uniformity. Although not necessary, 
this assumption will help simplifying our
analytical derivations in the remainder of this work, starting with Sec. \ref{sec:phasespace} 
(cf. \ref{app:linear} for more details). 

\begin{figure}[t]
\centerline{\resizebox{\textwidth}{!}{\includegraphics{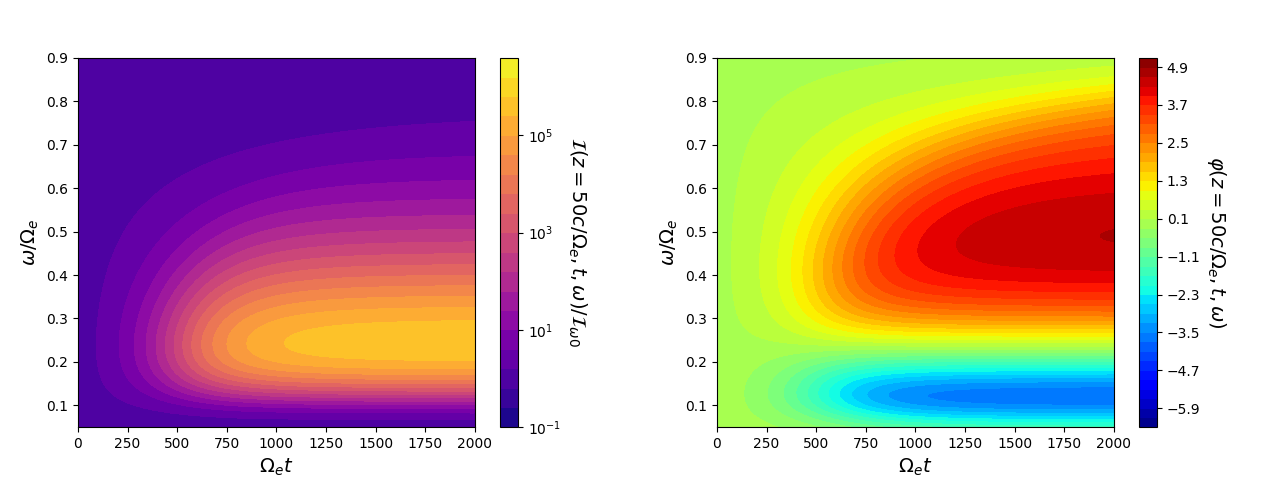}}} \
\vspace*{-2em}\newline\noindent(a) \hspace*{0.48\linewidth} (b)
\caption{Contour plots of $I (z, t, \omega)/I_{\omega 0}$ (a) and $\varphi (z, t, \omega)$ (b) are shown
for the linear evolution of the whistler wave packet and the same normalized parameters of Fig. \ref{fig:WGamma}. Here $I_{\omega 0} = {\rm const}$ 
and $\varphi_{\omega 0} = 0$ have been assumed  together with $z \Omega_e/c = 50$.}
\label{fig:IIphi}
\end{figure}

With the knowledge of $W(z, t, \omega)$ and $\Gamma(z, t, \omega)$, the nonlinear evolution of the chorus spectrum can be 
derived from the integration of Eqs. (\ref{eq:actionevolve}) and (\ref{eq:phaseevolve}) along the characteristics, recalling that
the right hand sides are, respectively, $2 \Gamma(z, t, \omega) I (z, t, \omega)$
and $- W(z, t, \omega)$ as noted above. Thus, solutions are formally written as
\begin{equation}
I (z, t, \omega) = I_{\omega 0} \left( T_\omega^{-1}(T_\omega(z)- t) \right) \exp \left( 2 \int_{T_\omega^{-1}(T_\omega(z)- t)}^z \frac{dz'}{v_{g \omega}(z')} \Gamma(z',t-T_\omega(z)+T_\omega(z'),\omega) \right) 
\; , \label{eq:Ikevolve}
\end{equation}
for the wave packet intensity $I (z, t, \omega)$, where 
\begin{equation}
T_\omega(z) \equiv \int_0^z  \frac{dz'}{v_{g \omega}(z')}  \; . \label{eq:Tktrans}
\end{equation}
Meanwhile, a similar solution can be written for phase shift $\varphi (z, t, \omega) \equiv \varphi_k (z, t)|_{k=K(z,\omega)}$
\begin{equation}
\varphi (z, t, \omega)  =  \varphi_{\omega 0} \left(T_\omega^{-1}(T_\omega(z)- t)\right) - \int_{T_\omega^{-1}(T_\omega(z)- t)}^z \frac{dz'}{v_{g \omega}(z')} W(z',t-T_\omega(z)+T_\omega(z'),\omega) \; . \label{eq:phikevolve}
\end{equation}
In the linear limit, where $W(z, t, \omega) = W(z, t=0, \omega)$ and $\Gamma(z, t, \omega) = \Gamma(z, t=0, \omega)$, Eqs. (\ref{eq:Ikevolve}) and (\ref{eq:phikevolve}) are readily computed and corresponding solutions are shown in the contour plots of Fig. \ref{fig:IIphi} for the same parameters of Fig. \ref{fig:WGamma} and $z \Omega_e/c = 50$.
Nonlinear evolution is all embedded in the time dependence of $W(z, t, \omega)$ and $\Gamma(z, t, \omega)$. 
In particular, we will show below that chorus chirping may be understood as the 
spectral frequency peak of $I (z, t, \omega)$ at a given spatial 
position shifting in time. Meanwhile, since the growth of the spectral peak is due to spontaneous emission of whistler
waves excited by hot electrons at the proper (instantaneous) wavelength and frequency,
chorus nonlinear evolution is, thus, clearly associated with maximization of wave particle power 
transfer \citep{Vomvoridis1982,Trakhtengerts2004,Omura2008}, as noted to be the case also for 
Alfv\'enic fluctuations in magnetized fusion plasmas \citep{Chen2016, Zonca2015}.
We will later come back to this very important point, with more insights and comments on the underlying physics. 
Summarizing, this analysis shows that chorus chirping rate can be predicted via analyzing $\partial_t \Gamma (z, t, \omega)$.
In particular, $\partial_t \Gamma$ can be derived from $\partial_t f_0$, that is from manipulation of the Dyson-like equation,
given in Sec. \ref{sec:phasespace}, as remarked in the Introduction. This derivation is carried out in Sec. \ref{sec:reddyson}, where we also show how 
$\partial_t \Gamma (z, t, \omega)$ and $\partial_t  W (z, t, \omega)$ are interlinked \citep{Zonca2017}.

\subsection{Phase Space Dynamics}
\label{sec:phasespace}

As shown in Sec. \ref{sec:waveequations}, hot electrons are localized about the equator and plasma non-uniformity effects are dominated
by the $\sim \zeta^4$ scaling of both $W(z, t, \omega)$ and $\Gamma(z, t, \omega)$. Thus, 
as noted in \ref{app:linear}, hot electrons can be 
approximated as a non-uniform source, characterized by the length scale $(A \xi)^{-1/2}$,  
neglecting magnetic field non-uniformity. 
This assumption helps simplifying our analytical derivations below and, thus, we choose to adopt it in the following. 
In order to simplify presentation, 
we also assume $\omega^2/\omega_p^2\ll 1$, so that expressions of $W(z, t, \omega)$ and $\Gamma(z, t, \omega)$
are reduced to
\begin{equation}
W(z, t, \omega)+i \Gamma(z, t, \omega) = \frac{n_e}{n} \zeta^4 \frac{\omega(\Omega_e-\omega)^2}{\Omega_e}
\left\langle \frac{1}{\Omega_e + k v_\parallel - \omega}
\left[ \frac{k v_\perp^2}{2 \omega} \frac{\partial}{\partial v_\parallel} -  \frac{\Omega_e}{\omega} \right] \hat f_0 \right\rangle \; , \label{eq:redWGamma}
\end{equation}
Here and in the following, $k=K(z=0,\omega)$ will always be assumed as obtained from the solution of the linear dispersion relation, $D_w=0$.
Meanwhile, as illustrated in \ref{app:linear}, we have integrated Eq. (\ref{eq:powerexchange}) 
by parts in $v_\perp$, extracted the expected hot electron non-uniformity 
scaling $\sim \zeta^4$, and denoted the response neglecting magnetic field non-uniformity
as $f_{0} \equiv n_e  \hat f_0$. 
Equation (\ref{eq:redWGamma}) suggests the connection of the phase shift and driving rate by a  localized hot electron source with 
that by a uniform hot electron source in uniform magnetic field. To make this more explicit, let us introduce the rescaled phase shift and driving rate,
$\bar W(z, t, \omega)$ and $\bar \Gamma(z, t, \omega)$, defined as
\begin{equation}
W(z, t, \omega)+i \Gamma(z, t, \omega) \equiv \zeta^4 \frac{\omega(\Omega_e-\omega)^2}{\Omega_e^2}
 \left( 1 - \frac{v_{r\omega}}{v_{g\omega}} \right)^{-2} \left[ \bar W(z, t, \omega)+i \bar \Gamma(z, t, \omega) \right]
 \; , \label{eq:scalWGamma}
\end{equation}
where $v_{r\omega} \equiv (\omega - \Omega_e)/k$ 
is the resonant velocity, and $v_{g\omega} = v_{g\omega}(z=0)$ is the group velocity at the equator, defined below Eq. (\ref{eq:WGamma}).
Thus, we rewrite Eq. (\ref{eq:redWGamma}) as
\begin{equation}
\bar W(z, t, \omega)+i \bar \Gamma(z, t, \omega) = \frac{n_e}{n} \left( 1 - \frac{v_{r\omega}}{v_{g\omega}} \right)^2
\left\langle \frac{\Omega_e}{\Omega_e + k v_\parallel - \omega} 
\left[ \frac{k v_\perp^2}{2 \omega} \frac{\partial}{\partial v_\parallel} -  \frac{\Omega_e}{\omega} \right] \hat f_0 \right\rangle \; . \label{eq:redWGammabar}
\end{equation}
The usefulness of introducing the factor $( 1 - v_{r\omega}/v_{g\omega} )^2$, 
where both resonant and group velocity are computed at $z=0$, will be
clarified in Sec. \ref{sec:reddyson}. There, we will also show that residual spatiotemporal dependences of $\hat f_0$, $\bar W$, and $\bar \Gamma$
will be via $t - z/v_{g\omega}$.

In the presence of a fluctuation spectrum in the form of Eq. (\ref{eq:choruswp}), the hot electrons distribution function, $f \equiv n_e  \hat f$ can be written as
\begin{equation}
\hat f (z,t)  =  \hat f_0 (z,t) + \frac{1}{2} \sum_k \left(  e^{i S_k(z,t) + i \alpha} \delta \hat{\bar{f}}_k (z,t) + c.c. \right) \; , 
\label{eq:elehatf}
\end{equation}
where $\hat f_0$, introduced in Eq. (\ref{eq:redWGamma}), denotes the $k=0$ component of $\hat f$;
and, for brevity, we have omitted the velocity space dependences of hot electron response. 
Similar to Eq. (\ref{eq:elef}), Equation (\ref{eq:elehatf}) follows from the fact that, given the chorus wave packet polarization 
properties discussed in Sec. \ref{sec:waveequations},
\begin{equation}
\frac{e}{m} \left( \delta \bar{\bm E}_k + \frac{ \bm v \times  \delta \bar{\bm B}_k }{c} \right) \cdot \frac{\partial}{\partial \bm v} = 
i \frac{e}{m} v_\perp \delta \bar E_k e^{i\alpha} \left[ \frac{k}{\omega}\frac{\partial}{\partial v_\parallel}
+ \left( 1 - \frac{k v_\parallel}{\omega} \right) \left( \frac{1}{v_\perp} \frac{\partial}{\partial v_\perp} + \frac{i}{v_\perp^2} \frac{\partial}{\partial \alpha} \right) \right]
\; . \label{eq:emforce}
\end{equation}
The evolution equation for $\hat f_0$ can be obtained from the Vlasov equation
\begin{eqnarray}
\left( \partial_t + v_\parallel \partial_z \right) \hat f_0 & = & \frac{1}{4} \sum_k  i \frac{e}{m} v_\perp \delta \bar E_k
\left[ \frac{k}{\omega}\frac{\partial}{\partial v_\parallel}
+ \left( 1 - \frac{k v_\parallel}{\omega} \right) \left( \frac{1}{v_\perp} \frac{\partial}{\partial v_\perp} + \frac{1}{v_\perp^2} \right) \right] \delta \hat{\bar{f}}_k^*
\nonumber \\ & & - \frac{1}{4} \sum_k  i \frac{e}{m} v_\perp \delta \bar E_k^*
\left[ \frac{k}{\omega}\frac{\partial}{\partial v_\parallel}
+ \left( 1 - \frac{k v_\parallel}{\omega} \right) \left( \frac{1}{v_\perp} \frac{\partial}{\partial v_\perp} + \frac{1}{v_\perp^2} \right) \right] \delta \hat{\bar{f}}_k
\; . \label{eq:barf0vla0}
\end{eqnarray}
Meanwhile, the fluctuating component of the hot electron response is given by
\begin{eqnarray}
{\cal L}_k\delta \hat{\bar{f}}_k & \equiv & \left[ k v_\parallel + \Omega_e - \omega -i \left( \partial_t + v_\parallel \partial_z \right) \right]\delta \hat{\bar{f}}_k
\nonumber \\ & &  =  \frac{e}{m} v_\perp \delta \bar E_k
\left[ \frac{k}{\omega}\frac{\partial}{\partial v_\parallel}
+ \left( 1 - \frac{k v_\parallel}{\omega} \right) \frac{1}{v_\perp} \frac{\partial}{\partial v_\perp} \right]  \hat f_0 \; . \label{eq:dbarfkvla0}
\end{eqnarray}
Here, ${\cal L}_k$ is a first order partial differential operator that can be ``formally inverted'' as propagator ${\cal L}_k^{-1}$,
which is an integral operator. Equation (\ref{eq:dbarfkvla0}) applies both linearly and nonlinearly; i.e., when $\hat f_0$ is
evolving in time due to resonant wave particle interactions. In particular, the action of the
$\left( \partial_t + v_\parallel \partial_z \right)$ operator on the phase modulation due to $\delta \bar E_k = |\delta \bar E_k| e^{i\varphi_k}$
must be computed noting that the hot electron induced frequency ($\Delta_\omega$) and 
wave number ($\Delta_k$) shifts due to an incremental change in the fluctuation spectrum 
still satisfy the whistler wave dispersion relation with good approximation. 
This assumption is based on observations
that chorus waves propagate in whistler mode; and it is consistent
with the interpretation of chorus chirping as ``whistler seeds'' that are excited in sequence and 
amplified by wave particle resonant interactions with hot electrons originally proposed by \citep{Omura2011}.
Thus, $\omega$ and $k$ in Eqs. (\ref{eq:barf0vla0}) and (\ref{eq:dbarfkvla0})
denote the wave packet frequency and wave number in the presence of the hot electron source 
and the finite amplitude chorus. The effect of the nonlinear frequency and wave number shifts due to 
an incremental change in the fluctuation spectrum is discussed in Sec. \ref{sec:choruschirp}.
Meanwhile, $\omega$ and $k$ in Eqs. (\ref{eq:barf0vla0}) and (\ref{eq:dbarfkvla0}) 
are interpreted as elements of the whistler fluctuation spectrum
that is considered dense (nearly continuous) and is self-consistently evolving in time as a whole in the presence
of the hot electron free energy source \citep{Zonca2017,Tao2020}, rather than considered as properly chosen ``whistler seeds'' that are
representative of the selected chorus element  \citep{Omura2011}.
This is one of the main differences of the present work with 
respect to the earlier analysis by \citep{Omura2011}, as discussed in the Introduction.  The other one consists in the analytic solution for the self-consistent nonlinear hot electron response in phase space \citep{Zonca2017,Tao2020}, which is discussed below. 
Section \ref{sec:choruschirp} also allows us to reconcile different interpretations of chorus
chirping \citep{Omura2011}) inside the same framework with a self-consistent
comprehensive vision and to address some of the issues
regarding sub-elements as presented in the recent work by  
\cite{Tsurutani2020}. 

When 
\begin{eqnarray}
\delta \hat{\bar{f}}_k =  {\cal L}_k^{-1} \left\{ \frac{e}{m} v_\perp \delta \bar E_k
\left[ \frac{k}{\omega}\frac{\partial}{\partial v_\parallel}
+ \left( 1 - \frac{k v_\parallel}{\omega} \right) \frac{1}{v_\perp} \frac{\partial}{\partial v_\perp} \right]  \hat f_0 \right\} \; , \label{eq:dbarfkvla1}
\end{eqnarray}
is substituted back into Eq. (\ref{eq:barf0vla0}), the ``formal solution'' for $\hat f_0$ is obtained and can be cast in 
the form of a Dyson-like equation \citep{Dyson1949, Schwinger1951, Itzykson80}
\begin{eqnarray}
\left( \partial_t + v_\parallel \partial_z \right) \hat f_0 & = & \frac{1}{4} \sum_k  i \frac{e}{m} v_\perp \delta \bar E_k
\left[ \frac{k}{\omega}\frac{\partial}{\partial v_\parallel}
+ \left( 1 - \frac{k v_\parallel}{\omega} \right) \left( \frac{1}{v_\perp} \frac{\partial}{\partial v_\perp} + \frac{1}{v_\perp^2} \right) \right] 
\nonumber \\ 
& &  \hspace*{2em}  \times  {\cal L}_k^{*-1} \left\{ \frac{e}{m} v_\perp \delta \bar E_k^*
\left[ \frac{k}{\omega}\frac{\partial}{\partial v_\parallel}
+ \left( 1 - \frac{k v_\parallel}{\omega} \right) \frac{1}{v_\perp} \frac{\partial}{\partial v_\perp} \right]  \hat f_0 \right\} \nonumber \\
& & - \frac{1}{4} \sum_k  i \frac{e}{m} v_\perp \delta \bar E_k^*
\left[ \frac{k}{\omega}\frac{\partial}{\partial v_\parallel}
+ \left( 1 - \frac{k v_\parallel}{\omega} \right) \left( \frac{1}{v_\perp} \frac{\partial}{\partial v_\perp} + \frac{1}{v_\perp^2} \right) \right] 
\nonumber \\ & & \hspace*{2em} \times  {\cal L}_k^{-1} \left\{ \frac{e}{m} v_\perp \delta \bar E_k
\left[ \frac{k}{\omega}\frac{\partial}{\partial v_\parallel}
+ \left( 1 - \frac{k v_\parallel}{\omega} \right) \frac{1}{v_\perp} \frac{\partial}{\partial v_\perp} \right]  \hat f_0 \right\}
\; . \label{eq:barf0vla1}
\end{eqnarray}
Connections of the present approach with the field theoretical description based on the Dyson-Schwinger
equations are extensively analyzed in Refs. \citep{Chen2016,Zonca2015} as far as magnetized fusion plasma
applications are concerned. Interested readers may find the same analyses specialized to chorus wave excitation
in \ref{app:dyson}. Here, we emphasize that the general theoretical framework \citep{Vanhove1954,Prigogine1962,Balescu1963,Altshul1966,Dupree1966,
Aamodt1967,Weinstock1969,Mima1973,Zonca2015, Zonca2015b, Chen2016, Zonca2017} is of crucial importance
for demonstrating that Eqs. (\ref{eq:dbarfkvla1}) and (\ref{eq:barf0vla1}) do indeed account for the 
phase space structures that determine the dominant nonlinear dynamics and phase space transport by chorus emission.
In fact, the dynamic description given by Eq. (\ref{eq:barf0vla1}) accounts for phase space nonlinear behaviors without fast temporal 
or spatial dependences, which correspond to the self-interaction of the fluctuation with the wavenumber of interest with itself. The resultant distortion of the 
hot electron distribution function, determined self-consistently in the presence of the finite amplitude fluctuation spectrum, constitutes
the ``renormalized'' hot electron response of interest for the present application.
Solving Eq. (\ref{eq:barf0vla1}) together with the wave equations,
Eqs. (\ref{eq:actionevolve}) and (\ref{eq:phaseevolve}), preserves the crucial underlying physics of chorus nonlinear evolution but 
is beyond the scope of the present work.
Here, we focus on chorus frequency chirping rather than on the details of phase space nonlinear dynamics and transport. Thus, in the next section 
we introduce a reduced (velocity space averaged) description of the Dyson-like equation that will allow us to derive nonlinear evolution equations for
$\bar W(z, t, \omega)$ and $\bar \Gamma(z, t, \omega)$ and, thereby, analytically address the dynamics of chorus chirping.

\section{Reduced Dyson-like equation}
\label{sec:reddyson}
  
Let us reconsider the simplified expressions of $\bar W(z, t, \omega)$ and 
$\bar \Gamma(z, t, \omega)$, Eq. (\ref{eq:redWGammabar}), 
obtained in Sec. \ref{sec:phasespace}. On the right hand side, formally 
consider $\hat f_0 \equiv$ \\ \noindent $\left( \partial_t + v_\parallel \partial_z \right)^{-1} \left( \partial_t + v_\parallel \partial_z \right) \hat f_0$ 
and use Eq. (\ref{eq:barf0vla1}) for the expression of $\left( \partial_t + v_\parallel \partial_z \right) \hat f_0$. 
In other words, we formally manipulate the Dyson-like equation obtained in Sec. \ref{sec:phasespace} 
and integrate in velocity space in order to obtain reduced expressions for time evolving
$\bar W(z, t, \omega)$ and $\bar \Gamma(z, t, \omega)$ rather than solving Eq. (\ref{eq:barf0vla1}) in the whole phase 
space \citep{Zonca2017, Tao2020}. This reduced approach becomes useful when the nonlinear particle response is dominated by 
resonant particles in the presence of a quasi-coherent (narrow) wave packet such as in the case of chorus. 
The same approach has been successfully applied to study energetic particle modes \citep{Zonca2015}
as well as the so called ``fishbone'' mode \citep{Chen2016} in fusion plasmas.
Let us also recall the approximation introduced at the beginning of Sec. \ref{sec:phasespace}, by which
we assume that hot electrons are a non-uniform source  localized about the equator, while the remaining dynamics 
is well described neglecting magnetic field non-uniformity. Thus, Eq. (\ref{eq:Tktrans})
gives $T_\omega(z) \simeq z/v_{g \omega}$ and $T^{-1}_\omega(t) \simeq v_{g \omega} t$. 
Furthermore, at any position $z$ sufficiently outside the  localized 
non-uniform hot electron source, $I (z,t,\omega)$ and $\varphi (z,t,\omega)$ are 
predominantly functions of $t - z/v_{g\omega}$, as can be verified from Eqs. (\ref{eq:Ikevolve}) and (\ref{eq:phikevolve})
computing $\partial_t$ and $\partial_z$ of those expressions.
Repeating the same argument, predominant dependence on $t - z/v_{g\omega}$ can be demonstrated for $\hat f_0$, $\bar W$ and $\bar \Gamma$. 
Residual $z$ dependences are neglected, since they account for magnetic field non-uniformity, which is omitted here
for simplicity, and modulation effects of the chorus wave packet due to the finite extent of the source region.  
These effects are reported in detailed numerical investigations
by \citep{Wu2020}, illustrating the role of magnetic field non-uniformity in breaking the symmetry between rising and falling 
tone chorus. In a more recent work \citep{Tao2021}, chorus nonlinear dynamics due to wave-particle interactions 
and magnetic field non-uniformity have been analyzed on the same footing within a newly developed phenomenological 
``TaRA model'', as anticipated in the Introduction. The present simplified theoretical description
reduces the dimensionality of the problem and allows us to adopt useful simplifications; 
\exgra, $(\partial_t + v_\parallel \partial_z ) \simeq (1 - v_{r\omega}/v_{g\omega}) \partial_t$ 
when dealing with resonant particles. Numerical studies of the complete Dyson-like Eq. (\ref{eq:barf0vla1}) will be given elsewhere.

Based on these assumptions, on the right hand side of Eq. (\ref{eq:barf0vla1}) we can write
\begin{equation}
i \left( \delta \bar E_k {\cal L}_k^{* -1}  \delta \bar E_k^* - \delta \bar E_k^* {\cal L}_k^{-1} \delta \bar E_k \right) =
i \left( \delta \bar E_k {\cal L}_k^{* -1} {\cal L}_k^{-1}  {\cal L}_k \delta \bar E_k^* - \delta \bar E_k^* {\cal L}_k^{-1} {\cal L}_k^{* -1} {\cal L}_k^{*} 
\delta \bar E_k \right) \; . \label{eq:forinv00}
\end{equation}
Noting 
\begin{equation}
{\cal L}_k^{* -1} {\cal L}_k^{-1} \simeq {\cal L}_k^{-1} {\cal L}_k^{* -1} \simeq \left[ {\cal L}_k {\cal L}_k^*\right]^{-1} = \left[ (\Omega_e + k v_\parallel - \omega)^2 
+  (1-v_{r\omega}/v_{g\omega})^2 \partial_t^2 \right]^{-1} \; , \label{eq:forinv0}
\end{equation}
where, again, the notation $[...]^{-1}$ always denotes the inverse of an operator within the square brackets,
Eq. (\ref{eq:forinv00}) can be cast as
\begin{eqnarray}
i \left( \delta \bar E_k {\cal L}_k^{* -1}  \delta \bar E_k^* - \delta \bar E_k^* {\cal L}_k^{-1} \delta \bar E_k \right) & = &
i \left\{ \delta \bar E_k \left[ (\Omega_e + k v_\parallel - \omega)^2 
+  (1-v_{r\omega}/v_{g\omega})^2 \partial_t^2 \right]^{-1}  {\cal L}_k \delta \bar E_k^* \right. \nonumber \\
& & \left. - \delta \bar E_k^* \left[ (\Omega_e + k v_\parallel - \omega)^2 
+  (1-v_{r\omega}/v_{g\omega})^2 \partial_t^2 \right]^{-1}  {\cal L}_k^{*} \delta \bar E_k \right\}
\nonumber \\
& \simeq & 2 |\delta \bar E_k| \left[ (\Omega_e + k v_\parallel - \omega)^2 + (1-v_{r\omega}/v_{g\omega})^2 \partial_t^2\right]^{-1} \nonumber \\
& & \times (1-v_{r\omega}/v_{g\omega}) \partial_t |\delta \bar E_k| \; . \label{eq:forinv1}
\end{eqnarray}
Here, as it can be verified by inspection, the operators action on the phase dependences in $\delta \bar E_k = |\delta \bar E_k| e^{i\varphi_k}$ and its
complex conjugate cancel each other and only $(1-v_{r\omega}/v_{g\omega}) \partial_t |\delta \bar E_k|$ survives. 
Using this expression and substituting Eq. (\ref{eq:barf0vla1}) back into Eq. (\ref{eq:redWGammabar}), 
we finally obtain, after tedious but straightforward algebra (cf. Appendix \ref{app:nonlinear})
\begin{eqnarray}
\bar W (\bar \omega) +i \bar \Gamma (\bar \omega) & =  & \frac{n_e}{n} \left( 1 - \frac{v_{r\bar\omega}}{v_{g\bar\omega}} \right)^2
\left\langle \frac{v_\perp^4}{2} \Omega_e \left[ \Omega_e + \bar k v_\parallel - \bar \omega - i (1-v_{r\bar\omega}/v_{g\bar\omega}) \partial_t\right]^{-1}   \right. \nonumber \\
& & \times \left( \frac{\bar k}{\bar \omega} \frac{\partial}{\partial v_\parallel} - \frac{2}{\left\langle v_\perp^2\right\rangle} \frac{\Omega_e}{\bar \omega} \right)  \left( \partial_t + v_\parallel \partial_z \right)^{-1} \sum_k  \frac{e^2}{2 m^2} |\delta \bar E_k| \frac{k}{\omega} \frac{\partial}{\partial v_\parallel} \nonumber \\
& &  \times \left[ (\Omega_e + k v_\parallel - \omega)^2 + (1-v_{r\omega}/v_{g\omega})^2 \partial_t^2\right]^{-1} (1-v_{r\omega}/v_{g\omega}) 
\nonumber \\ & &  \times \left. \partial_t |\delta \bar E_k| \left( \frac{k}{\omega} \frac{\partial}{\partial v_\parallel} - \frac{2}{\left\langle v_\perp^2\right\rangle} \frac{\Omega_e}{\omega} \right) \hat f_0 \right\rangle \; . \label{eq:redWGammabar0}
\end{eqnarray}
Here, we have repeatedly integrated by parts in $v_\perp$, from outside to the inside, in order to remove any $\partial/\partial v_\perp$ in the final expression. Moreover, 
for simplicity of notation, we have explicitly indicated only the frequency dependence of $\bar W$ and $\bar\Gamma$, leaving implicit the dependence on 
$t - z/v_{g\bar\omega}$. Finally $\left\langle v_\perp^2\right\rangle \equiv \left\langle v_\perp^2 \hat f_0\right\rangle/\left\langle \hat f_0\right\rangle$, 
and we have denoted the current frequency and wave number satisfying the lowest order whistler wave dispersion relation 
as $\bar \omega$ and $\bar k$ in order to distinguish them from $\omega$ and $k$ in the running summation over the fluctuation spectrum. 

Equation (\ref{eq:redWGammabar0}) still contains all the information embedded in the solution of the Dyson-like equation, Eq. (\ref{eq:barf0vla1}), 
via complicated integro-differential operators. In order to make further progress, we explicitly carry out the velocity space integration adopting 
two assumptions: (i) the chorus spectrum is narrow, such that $(\omega, k) \simeq (\bar\omega, \bar k)$ and $\partial_t 
|\delta \bar E_k| \simeq \partial_t |\delta \bar E_{\bar k}|$; (ii) chorus chirping is due to the subsequent emission of different waves 
belonging to the whistler wave continuum, which are excited in turns to maximize wave particle power transfer. 
The assumption (ii) was already introduced in the remarks following 
Eq. (\ref{eq:barf0vla1}) in Sec. \ref{sec:phasespace} and will be further discussed below in Sec. \ref{sec:choruschirp}.
Meanwhile, both assumptions are based on the chorus spectral features and are the same as
those of fluctuation spectra in the aforementioned fusion applications \citep{Chen2016, Zonca2015}.
After tedious but straightforward algebra, some details of which are reported in Appendix \ref{app:nonlinear2} for interested readers,
real and imaginary parts of Eq. (\ref{eq:redWGammabar0}) can be cast as
\begin{eqnarray}
\Omega_e^{-1} \partial_t \bar W (\bar \omega) & =  & \left[ \Omega_e \frac{\partial}{\partial \bar \omega} - \frac{2 \Omega_e^2}{\bar k^2 \left\langle v_\perp^2\right\rangle} 
\left( 1 - \frac{v_{r\bar\omega}}{v_{g\bar\omega}} \right) \right] \Omega_e \frac{\partial}{\partial \bar \omega} \nonumber \\
& & \times \sum_k \frac{\left( \omega - \bar \omega \right)}{2 \Omega_e} \left[ \frac{\left( \omega - \bar \omega \right)^2}{4 \Omega_e^2}+\Omega_e^{-2} \partial_t^2\right]^{-1} \nonumber \\
& &  \times \frac{\left\langle\left\langle \omega_{{\rm tr}k}^4 \right\rangle\right\rangle}{4 \Omega_e^4 (1-v_{r\omega}/v_{g\omega})^4}
\left( \frac{\bar \Gamma (\omega) + \bar \Gamma (\bar \omega)}{2} \right) \; ; \label{eq:redWbar0}
\end{eqnarray}
and, denoting as $\bar \Gamma_L$ the linear (initial) normalized hot electron driving rate, 
\begin{eqnarray}
\bar \Gamma (\bar \omega) - \bar \Gamma_L (\bar \omega) & =  & \left[ \Omega_e \frac{\partial}{\partial \bar \omega} - \frac{2 \Omega_e^2}{\bar k^2 \left\langle v_\perp^2\right\rangle} 
\left( 1 - \frac{v_{r\bar\omega}}{v_{g\bar\omega}} \right) \right] \Omega_e \frac{\partial}{\partial \bar \omega} \nonumber \\
& & \times \sum_k \left[ \frac{\left( \omega - \bar \omega \right)^2}{4 \Omega_e^2}+\Omega_e^{-2} \partial_t^2\right]^{-1}  
\nonumber \\ & &  
\times \frac{\left\langle\left\langle \omega_{{\rm tr}k}^4 \right\rangle\right\rangle}{4 \Omega_e^4 (1-v_{r\omega}/v_{g\omega})^4}
\left( \frac{\bar \Gamma (\omega) + \bar \Gamma (\bar \omega)}{2} \right) \; , \label{eq:redGammabar0}
\end{eqnarray}
where we have introduced the wave particle trapping frequency definition
\begin{equation}
\left\langle\left\langle \omega_{{\rm tr}k}^4 \right\rangle\right\rangle \equiv \left \langle v_\perp^2 \omega_{{\rm tr}k}^4 \hat f_0 \right\rangle/\left \langle v_\perp^2 \hat f_0\right\rangle
\; , \label{eq:wptrapomega}
\end{equation}
with $\omega_{{\rm tr}k}^2 = |(e/m) k^2 v_\perp \delta \bar E_k/\omega|$. 

Equations (\ref{eq:redWbar0}) and (\ref{eq:redGammabar0}) are still complicated nonlinear integro-differential equations, 
but they have been significantly simplified (or reduced) with respect to the original Dyson-like equation, Eq. (\ref{eq:barf0vla1}). 
These equations are the primary theoretical results of the present work and show that $\bar W$ and 
$\bar \Gamma$ evolution equations are interlinked, as expected and as anticipated in Sec. \ref{sec:waveequations}. They describe
a variety of nonlinear dynamics, including chorus chirping and modulation of the chorus wave packets 
on a time scale $\sim \left\langle\left\langle \omega_{{\rm tr}k}^4 \right\rangle\right\rangle^{-1/4}$. To see
this more clearly, let us introduce the optimal ordering for Eqs. (\ref{eq:redWbar0}) and (\ref{eq:redGammabar0}).
The width of the fluctuation spectrum can be estimated as 
\begin{equation}
\frac{\left| \omega - \bar \omega \right|}{2 \Omega_e} \sim \frac{ \partial_t \bar W}{ \Omega \bar \Gamma_{NL}} \; . \label{eq:specwidth}
\end{equation}
Meanwhile, assuming $|\partial_{\bar \omega}^2 \Gamma_{NL}| \gaeq |\partial_{\bar \omega}^2 \Gamma_L|$, ordering all terms of Eq. (\ref{eq:redGammabar0}) on the same footing gives 
\begin{equation}
|\partial_t| \sim \left| \omega - \bar \omega \right| \sim |\partial_{\bar \omega}|^{-1} \sim \left\langle\left\langle \omega_{{\rm tr}k}^4 \right\rangle\right\rangle^{1/4} \sim \hat \gamma_e \; , \label{eq:optimord}
\end{equation}
where $\hat \gamma_e$ is the peak value of the linear hot electron driving rate at the equator. Equation (\ref{eq:optimord}) describes a whistler wave packet that grows and saturates locally due to wave particle trapping. However, if chirping consistent with Eq. (\ref{eq:1}) sets in as in chorus spontaneous emission, saturation at the level
of Eq. (\ref{eq:optimord}) is not possible and the wave packet can grow further. Equation (\ref{eq:redGammabar0}), then, 
suggests
\begin{equation}
|\partial_t| \sim \left\langle\left\langle \omega_{{\rm tr}k}^4 \right\rangle\right\rangle^{1/2} |\partial_{\bar \omega}| 
\sim \frac{\left\langle\left\langle\omega_{{\rm tr}k}^4 \right\rangle\right\rangle^{1/2}}{\left| \omega - \bar \omega \right|} \gaeq  \left| \omega - \bar \omega \right| \; . \label{eq:chirpord}
\end{equation}
This ordering corresponds to a characteristic nonlinear time, $\tau_{NL} \sim |\partial_t|^{-1} \sim  \Gamma_{NL}^{-1}$ that is shorter than
 the wave particle auto-correlation time, $|\Delta \omega|^{-1} \sim \left| \omega - \bar \omega \right|^{-1}$. 
 In particular, this ordering is consistent with chorus chirping, and Eq. (\ref{eq:redGammabar0}) readily yields Eq. (\ref{eq:1})
when we assume a quasi-coherent (nearly monochromatic) fluctuation spectrum. In fact, keeping the leading terms  only in  Eq. (\ref{eq:redGammabar0}), consistent with Eq. (\ref{eq:chirpord}), we have 
that $\Omega_e^{-2} \partial_t^2$ dominates in the integral
operator definition, Eq. (\ref{eq:redGammabar0}), and 
\begin{eqnarray}
\frac{\partial^2}{\partial t^2} \bar \Gamma_{NL} (\bar \omega) \simeq \left( \sum_k  \frac{\left\langle\left\langle \omega_{{\rm tr}k}^4 \right\rangle\right\rangle}{4 (1-v_{r\omega}/v_{g\omega})^4} \right) \frac{\partial^2}{\partial \bar \omega^2} \bar \Gamma_{NL} (\bar \omega)  . \label{eq:redGammabar1}
\end{eqnarray}
This result suggest that there exists a self similar solution for $\bar \Gamma_{NL}$ that ballistically propagates in $\omega$-space at a rate given by the square root of the 
quantity in square parentheses on the right hand side. This ballistic propagation corresponds to the analogous ballistic propagation of hot electron phase space structures, described by Eq. (\ref{eq:barf0vla1}); and it is in one-on-one correspondence
with the analogous ballistic propagation of phase space zonal structures connected
with energetic particle avalanches in fusion plasmas \citep{Chen2016, Zonca2015}. Details obviously depend on the actual form of the spectrum, but it is readily verified that, for 
nearly monochromatic chorus element, Eq. (\ref{eq:redGammabar1}) yields
\begin{equation}
\frac{\partial \omega}{\partial t} = \pm \frac{1}{2}  \frac{\left\langle\left\langle \omega_{{\rm tr}k}^4 \right\rangle\right\rangle^{1/2}}{(1-v_{r\omega}/v_{g\omega})^2} \; . \label{eq:choruschirpboth}
\end{equation}
Thus, the present theoretical framework is consistent with both upward and downward frequency sweeping of chorus structures and, thus,
consistent with the recent work by \cite{Wu2020}. 
This is the first theoretical prediction for the downward chirping of parallel propagating waves, as far as we are aware of.
However, it should also be noted that statistical observations of chorus falling tones show that they are mainly very oblique rather than parallel
\citep{Li2011}, and that \citet{Mourenas2015} have predicted the possible existence of both positive and negative frequency chirping 
for such very oblique chorus waves based on the maximization of the nonlinear growth rate of \citet{Shklyar2009}.
More detailed discussions on the chirping direction will be given below in Sec. \ref{sec:choruschirp}.
Focusing, here, on the positive sign of Eq. (\ref{eq:choruschirpboth}), by direct inspection it can be noted that
it coincides with Eq. (\ref{eq:1}) for $R=1/2$ as anticipated in the Introduction \citep{Vomvoridis1982}. Equation (\ref{eq:choruschirpboth})
improves an earlier estimate by the same authors \citep{Zonca2017} and, to our knowledge, is
the first self-consistent analytical demonstration of the conjecture by \cite{Vomvoridis1982} in its exact initial formulation. 
More generally, Eq. (\ref{eq:redGammabar1})
suggests why the chorus chirping rate is not always given by $R=1/2$, the limiting case for a nearly monochromatic spectrum, but may
vary depending on the excitation conditions; \exgra, the initial hot electron distribution function.

\section{On chorus chirping}
\label{sec:choruschirp}

In order to further illuminate the features of chorus dynamics as predicted by Eqs. (\ref{eq:redWbar0}) and (\ref{eq:redGammabar0}) , we have solved
them numerically, together with the wave equations of Sec. \ref{sec:waveequations}, using a 4th order Runge-Kutta method. 
To exploit the dense nature of the whistler wave spectrum, we introduce the dimensionless intensity ${\cal I} (\omega)$ such that 
nonlinearity effects become important when ${\cal I} (\omega) \sim {\cal O}(1)$ (cf. \ref{app:numsol}). 

\begin{figure}[t]
	\begin{center}
		\resizebox{\textwidth}{!}{\includegraphics{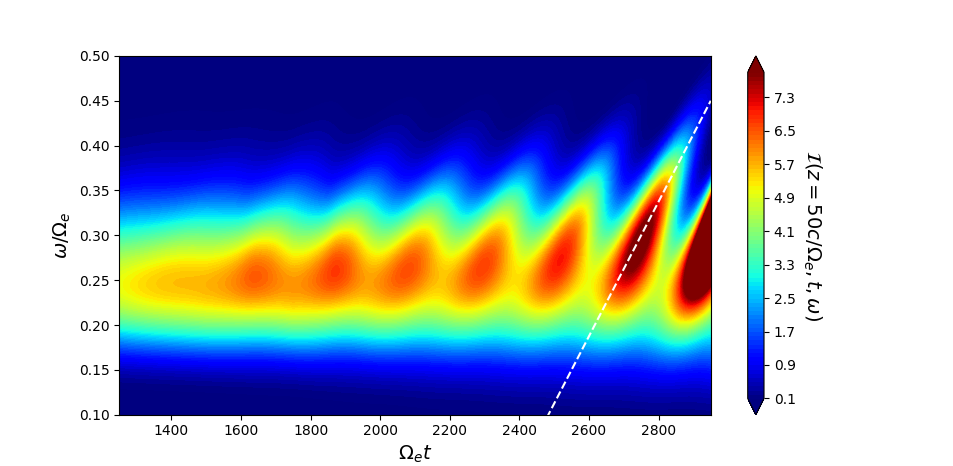}}
	\end{center}
\caption{Contour plot of the nonlinear evolution of ${\cal I} (\omega)$ for a uniform source ${\cal S} = 2 \times 10^{-5}$ in Eq. (\ref{eq:ievolve}). 
The position is fixed at $z \Omega_e/c = 50$ and parameters are the same as in Fig. \ref{fig:WGamma}. The white dashed line passing through the
chorus element beginning at $\Omega_e t \sim 2600$ represents the average chirping rate $\partial_t \omega_0 = 7.5 \times 10^{-4} \Omega_e^2$.}
\label{fig:NLspectrum}
\end{figure}

\begin{figure}[t]
	\begin{center}
		\resizebox{\textwidth}{!}{\includegraphics{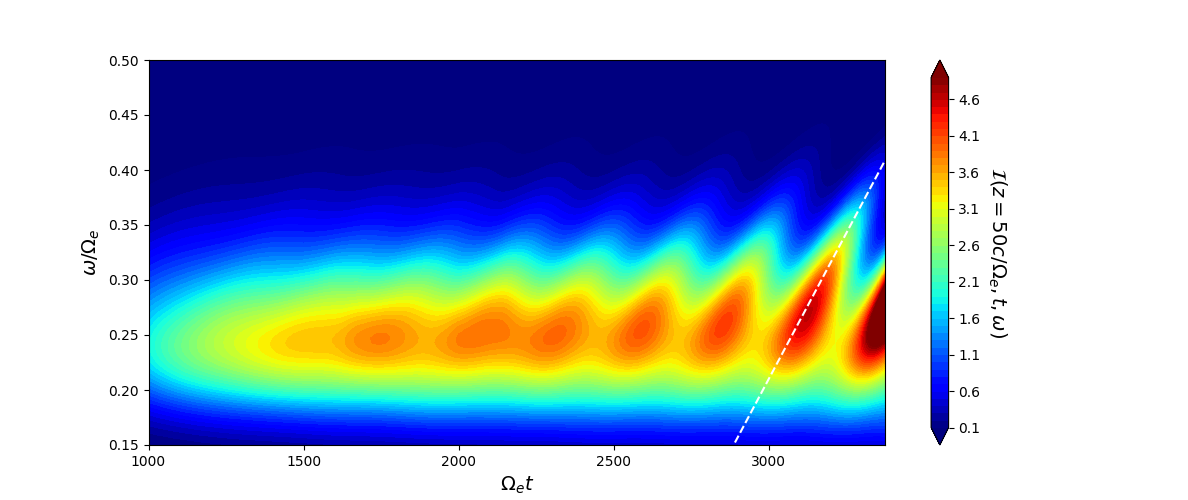}}
	\end{center}
\caption{Contour plot of the nonlinear evolution of ${\cal I} (\omega)$ for a uniform source ${\cal S} = \sqrt{2} \times 10^{-5}$ in Eq. (\ref{eq:ievolve}). 
As in Fig. \ref{fig:NLspectrum}, the position is fixed at $z \Omega_e/c = 50$ and parameters are the same as in Fig. \ref{fig:WGamma}. The white dashed line passing through the
chorus element beginning at $\Omega_e t \sim 3000$ represents the average chirping rate $\partial_t \omega_0 = 5.3 \times 10^{-4} \Omega_e^2$ as obtained
from PIC simulation by the DAWN code in \citep{Tao2017a}.}
\label{fig:NLspectrum-b}
\end{figure}

Assuming fixed $z \Omega_e/c = 50$ and parameters as in Fig. \ref{fig:WGamma}, the nonlinear evolution
of ${\cal I} (z = 50 c/\Omega_e, t, \omega)$ is shown in Fig. \ref{fig:NLspectrum}. Here, rather than assuming
a specific form of the initial spectrum, we assumed vanishing initial conditions and a constant slow external stirring
that, in the absence of supra-thermal electrons, would give an intensity spectrum ${\cal I} = {\cal S}^2\Omega_e^2 t^2$,
corresponding to $|\delta \bar E_k|$ linearly increasing with time (cf. \ref{app:numsol}). In Fig. \ref{fig:NLspectrum}, the source strength is
${\cal S} = 2 \times 10^{-5}$. Furthermore, we use 
a discretization in $\omega$ space with $221$ grid points in the interval $\omega/\Omega_e \in [0.05, 0.9]$ and adopt a 
Savitzky--Golay filter fitting  sub-sets of 19 adjacent data points with a fourth order degree polynomial to  
ensure regularity of the derivatives in $\omega$-space. A fourth order Runge-Kutta integration in time is adopted
with variable time step, gradually decreasing from an initial $\Omega_e \Delta t = 1.25 \times 10^{-1}$ in the early linear evolution to 
$\Omega_e \Delta t = 3.125 \times 10^{-2}$ in the later nonlinear phase at $\Omega_e t > 1750$. This choice ensures that Courant condition
is well satisfied. The routine solving Eqs. (\ref{eq:Gdefs}) to (\ref{eq:phievolve}), closed by Eqs. (\ref{eq:WGammarenorm}) 
and (\ref{eq:Iphirenorm}) together with the aforementioned boundary 
conditions, is written in Python and uses Python standard libraries. In order to illustrate the robustness of the present numerical results
as parameters are varied, Fig. \ref{fig:NLspectrum-b} shows the nonlinear evolution
of ${\cal I} (z = 50 c/\Omega_e, t, \omega)$  for ${\cal S} = \sqrt{2} \times 10^{-5}$ and same physical parameters of Fig. \ref{fig:NLspectrum}. In this case, the 
Savitzky--Golay filter is reduced to fitting sub-sets of 15 adjacent data points with a fourth order degree polynomial. 

\begin{figure}[t]
	\centerline{\resizebox{\textwidth}{!}{\includegraphics{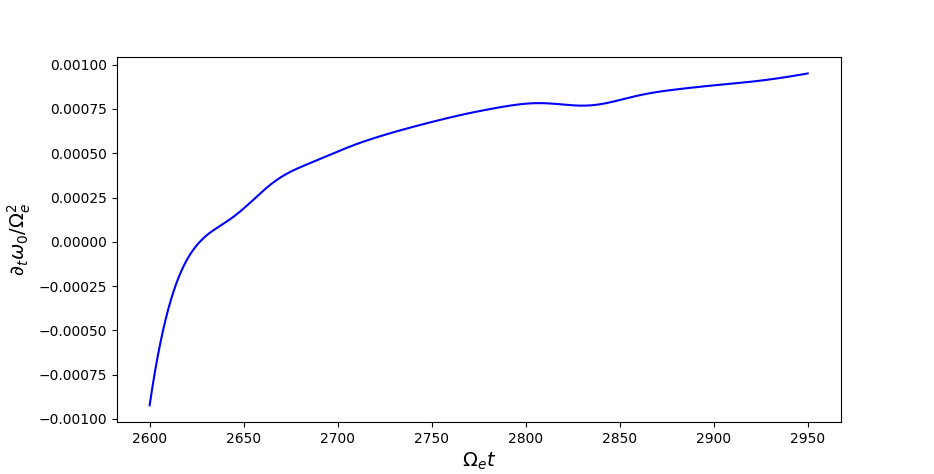}}}
\caption{Instantaneous chirping rate of the intensity peak of the
rising tone chorus element beginning at $\Omega_e t \sim 2600$ in Fig. \ref{fig:NLspectrum}.}
\label{fig:chirp}
\end{figure}

\begin{figure}[t]
	\centerline{\resizebox{\textwidth}{!}{\includegraphics{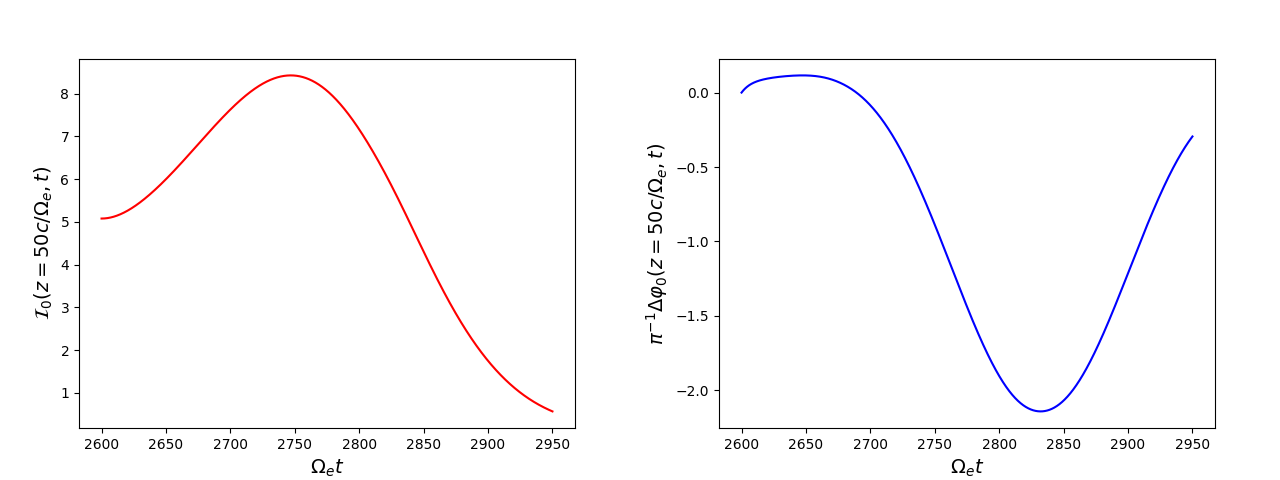}}}
	\vspace*{-.5em} (a) \hspace*{0.46\textwidth} (b)
\caption{Time evolution of intensity peak ${\cal I}_0 (z = 50 c/\Omega_e, t)$ (a) and corresponding phase shift $\Delta \varphi_0 (z = 50 c/\Omega_e, t)$ (b) 
for the rising tone chorus element beginning at $\Omega_e t \sim 2600$ in Fig. \ref{fig:NLspectrum}.} 
\label{fig:chirpel}
\end{figure}

\begin{figure}[t]
	\centerline{\resizebox{\textwidth}{!}{\includegraphics{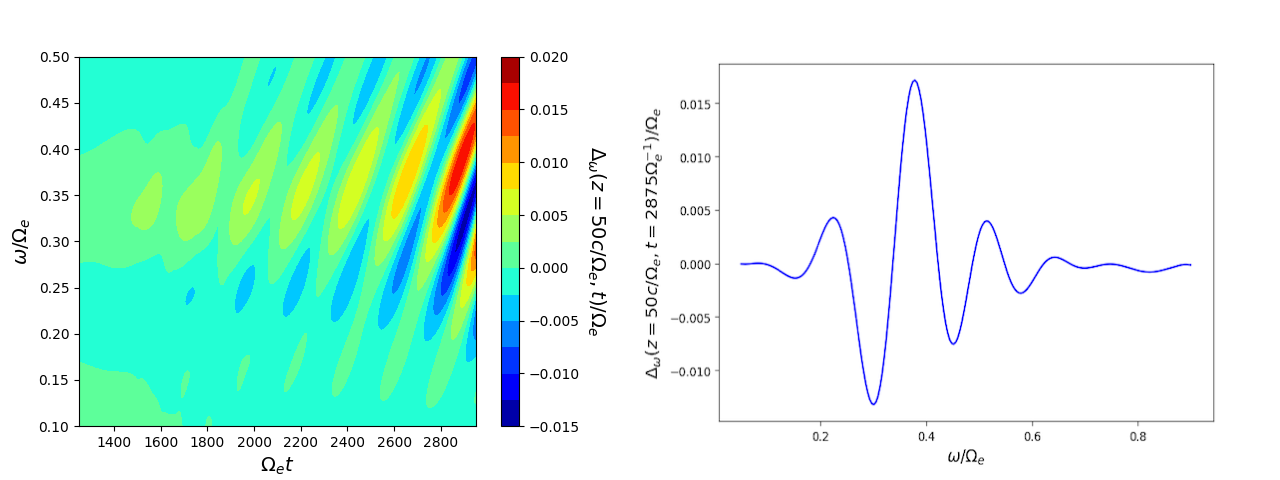}}}
	\vspace*{-.5em} (a) \hspace*{0.46\textwidth} (b)
\caption{Contour plot of the nonlinear frequency shift, $\Delta_\omega (z = 50 c/\Omega_e, t)$ (a); and  
snapshot of $\Delta_\omega (z = 50 c/\Omega_e, t = 2875/\Omega_e)$ (a).} 
\label{fig:deltaw}
\end{figure}

\begin{figure}[t]
		\centerline{\resizebox{\textwidth}{!}{\includegraphics{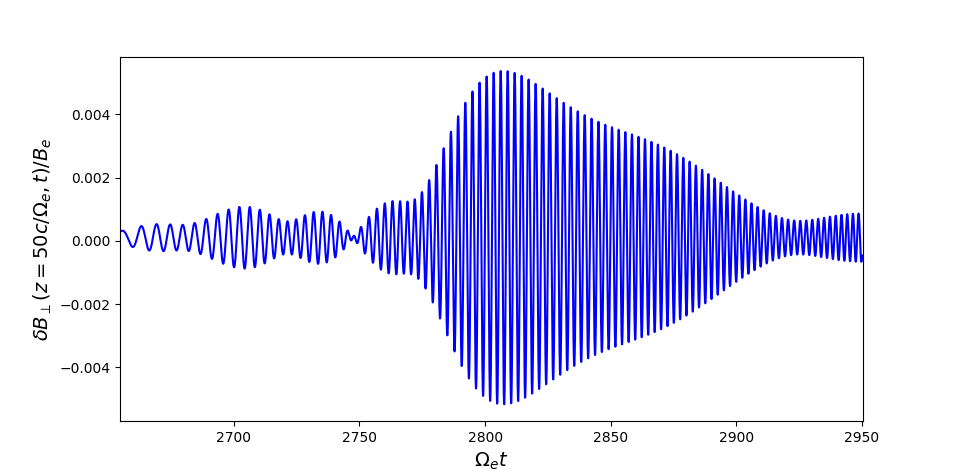}}}
\caption{Temporal structure of the perpendicular magnetic field fluctuation, $| \delta \bm B_\perp|/B_e$
for the rising tone chorus element beginning at $\Omega_e t \sim 2600$ in Fig. \ref{fig:NLspectrum}. Formation of 
``subpackets'' or ``subelements'' \citep{Santolik2003} is clearly illustrated.} 
\label{fig:deltaBw}
\end{figure}

Various distinctive features clearly emerge in Fig. \ref{fig:NLspectrum} and Fig. \ref{fig:NLspectrum-b}. After the initial formation of the ``linear'' fluctuation spectrum,
clear modulations at the $\left\langle\left\langle \omega_{\rm{tr}k}^4\right\rangle\right\rangle^{1/4}$ frequency become increasingly more evident
as the fluctuation intensity $\cal I$ grows larger than unity, as expected from the previous theoretical analysis and from \citep{Tao2017a}. 
Note that these modulations are different from the amplitude modulations within one chorus element leading to the so-called 
``subpackets'' or ``subelements'' \citep{Santolik2003}. However, they stem from the same physics; that is, the spectrum intensity modulation due to 
the finite frequency width of the wave packet as shown in Eqs. (\ref{eq:Gdefs}).
Nonlinear oscillations, as intensity increases,
are accompanied by gradually increasing frequency chirping, which can be both up or down.
This behavior is consistent with Eq. (\ref{eq:choruschirpboth}) and, despite no clear falling tone chorus element is observed here, it is also consistent with the recent numerical
investigation by \citep{Wu2020}.
Further strengthening of the nonlinear oscillations due to the continuous energy injection in the system by the uniform source ${\cal S}$, which
is amplified via resonant wave particle power exchange and structure formation in the phase space, breaks the up-down symmetry in the chirping process because of the lack of symmetry (in frequency) of the linear drive about its maximum (cf. Fig. \ref{fig:WGamma}) and because 
of the symmetry breaking term in the first line on the right hand side of Eq. (\ref{eq:redGammabar2}). Another origin of symmetry breaking
in frequency chirping is due to the non-uniformity due to the ambient magnetic field \citep{Wu2020}, which, however, is 
neglected for the sake of simplicity in the present theoretical analysis. 
Focusing on the rising tone chorus element beginning at $\Omega_e t \sim 2600$ in Fig. \ref{fig:NLspectrum}, the frequency chirping is well represented by 
Eq. (\ref{eq:choruschirpboth}) and is fitted by the average chirping rate $\partial_t \omega_0 = 7.5 \times 10^{-4} \Omega_e^2$. For the somewhat
weaker power injection in Fig. \ref{fig:NLspectrum-b}, the chirping of rising tone chorus element beginning at $\Omega_e t \sim 3000$ 
agrees remarkably well with the average chirping rate $\partial_t \omega_0 = 5.3 \times 10^{-4} \Omega_e^2$ as obtained
from PIC simulation by the DAWN code in \citep{Tao2017a} and, again, is visually given by the white dashed line. 
The average chirping rate dependence on the fluctuation intensity further confirms Eq. (\ref{eq:choruschirpboth}). The average chirping rate
is also confirmed by the instantaneous chirping rate of the intensity peak given in Fig. \ref{fig:chirp}. Noting $\partial_t \omega_0/\Omega_e^2$ is starting from negative values,
as noted above, is consistent with the possibility of both up- and down-chirping and, thus, with Eq. (\ref{eq:choruschirpboth}). However,
here, we cannot observe a clear formation of a falling tone chorus element unlike in \citep{Wu2020}, despite the evidence of initial
down-chirping. 
As the rising tone chorus element is 
clearly formed with the corresponding phase space structure, the chirping rate reaches up to its average value as visually suggested by the white dashed line
in Fig. \ref{fig:NLspectrum}. 
Time evolution of intensity peak ${\cal I}_0 (z = 50 c/\Omega_e, t)$ and
corresponding phase shift $\Delta \varphi_0 (z = 50 c/\Omega_e, t)$ for this chorus element are given in Fig. \ref{fig:chirpel} and further clarify the underlying physics. 
An important conclusion we may draw from Fig. \ref{fig:chirpel} is that intensity grows while
$\Delta \varphi_0 \simeq 0$; i.e., during phase locking. This behavior is due to phase bunching of
both trapped and untrapped resonant particles, which most effectively drive the chorus wave-packet.
The same behavior allows us drawing strong connection with the analogous behavior of energetic
particle avalanches in fusion plasmas \citep{Zonca2015,Chen2016}. Meanwhile,
the intensity peak takes place when $\Delta \varphi_0 = \pi$ and resonant particles phase locking is lost yielding the end
of the chorus event.
This mechanism can be viewed as the chorus wave packet slipping over the population of resonant electrons maximizing wave particle power extraction, and suggests the analogy with super-radiance in free electron lasers introduced by \citep{Zonca2015,Chen2016} with regard to energetic particle mode convective amplification in fusion plasmas.  Analogies with the free electron laser 
were also noted by numerical simulation studies in \citep{SotoChavez2012}.
That it is indeed maximization of  wave particle power transfer \citep{Vomvoridis1982,Omura2008,Zonca2015,Zonca2015b,Chen2016} 
that dictates the nonlinear chorus dynamics and frequency chirping
is further demonstrated by Fig. \ref{fig:deltaw}. Figure \ref{fig:deltaw} (a) shows that the nonlinear frequency shift, $\Delta_\omega (z = 50 c/\Omega_e, t)$,  
remains small during the whole nonlinear evolution and, in particular, much smaller than the dynamic range of frequency chirping,
consistent with the assumption that each elementary wave constituting the
chorus wave packet satisfies the whistler wave dispersion relation at the lowest order. Figure \ref{fig:deltaw} (b), meanwhile, 
shows a snapshot of $\Delta_\omega (z = 50 c/\Omega_e, t = 2875/\Omega_e)$.
By definition, at the intensity peak the wave particle power transfer is maximized; and, since the chirping process is 
spontaneously triggered by the underlying instability, the nonlinear evolution follows the 
maximum possible intensity growth or minimum possible intensity decrease. In fact, it is important to recognize that
power transfer is maximized even in the intensity decreasing phase \citep{Zonca2015,Zonca2015b,Chen2016} .
Another aspect that is clarified by the present approach and is the issue of ``subpackets'' or ``subelements'' \citep{Santolik2003} formation within
a single chorus element. While Fig. \ref{fig:NLspectrum} and Fig. \ref{fig:NLspectrum-b} display the spectrum intensity only, Fig. \ref{fig:deltaBw}
illustrates the temporal structure of the perpendicular magnetic field fluctuation, $| \delta \bm B_\perp|/B_e$, reconstructed
from Eqs. (\ref{eq:choruswp}) and (\ref{eq:ampphase}), 
for the rising tone chorus element beginning at $\Omega_e t \sim 2600$ in Fig. \ref{fig:NLspectrum}. The formation of
subelements is clear, qualitatively and quantitatively consistent with the PIC simulation by the DAWN code done with the same parameters in \citep{Tao2017a}.
This evidence supports the corresponding original interpretation provided therein that chorus sub-element formation is to be attributed to 
the phase modulation and ``self-consistent evolution of resonant particle phase-space structures and spatiotemporal features 
of the fluctuation spectrum'', proposed by \citep{ONeil1965} when analyzing collisionless damping of nonlinear plasma oscillations.
These results also clarify that nonlinear oscillations are connected with the width of the fluctuation intensity spectrum and
stem from the same underlying physics, as noted above. In close connection and consistent with the present analysis, it is important to quote
the recent statistical results from \citet{Zhang2020a} on observed typical wave packet lengths, amplitudes, and frequency variations of rising tone chorus elements. 
Short packets have been explained by \citet{Zhang2020a} and \citet{Nunn2021} as resulting from trapping-related amplitude modulations 
for packets longer than about 10 wave periods, and as a result of wave superposition of two well-separated waves 
sensibly farther than a trapping period for shorter packets.
Formation of subpackets in chorus emission was also recently 
analyzed by \citep{Hanzelka2020} adopting the sequential triggering model by \cite{Omura2011}.

\begin{figure}[t]
		\centerline{\resizebox{\textwidth}{!}{\includegraphics{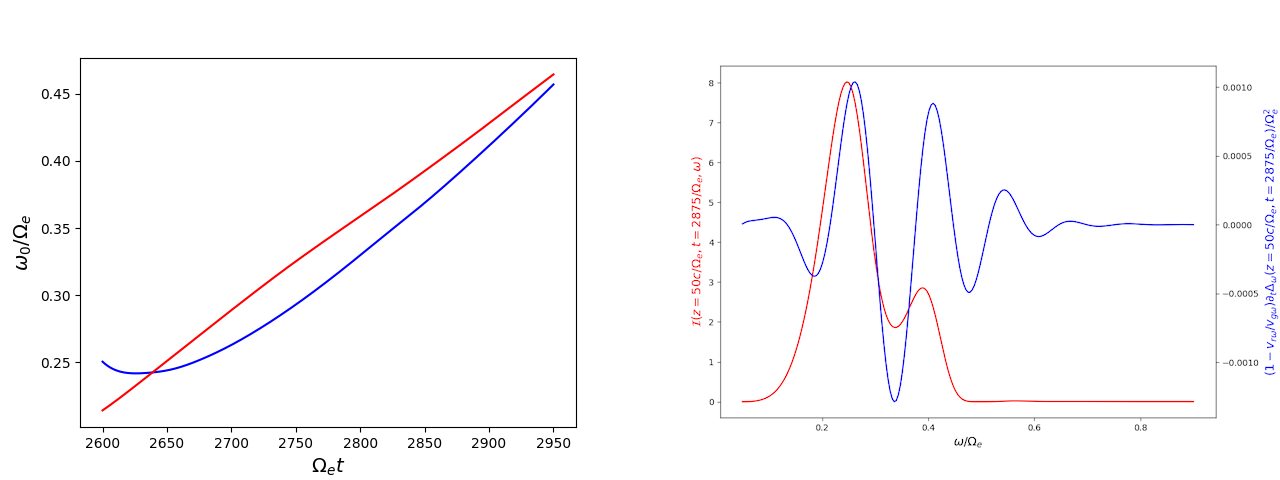}}}
	\vspace*{-.5em} (a) \hspace*{0.46\textwidth} (b)
\caption{(a) Time evolution of intensity peak frequency (blue line) and of the frequency
of the peak of the corresponding maximum in the rate of change of nonlinear frequency shift (red line)
for the rising tone chorus element beginning considered in Fig. \ref{fig:NLspectrum}. (b) Snapshot at $\Omega_e t =2875$
of the fluctuation intensity and of the $\partial_t \omega_{\rm res}$ as a function 
of frequency.} 
\label{fig:omura}
\end{figure}

Given the present theoretical analysis and numerical solutions, the explanation of chorus frequency chirping given by 
 \citep{Omura2011} may seemingly be in contrast with the present results. As anticipated in the Introduction, the reason for frequency chirping 
was explained as  due to the nonlinear current parallel to the wave magnetic field ($J_B$), 
which causes a nonlinear frequency shift. More precisely, the physics mechanism underlying chirping  
is the sequence of ``whistler seeds'' that are excited and 
amplified by wave particle resonant interactions with supra-thermal electrons. In the present
work, the fluctuation spectrum is self-consistently evolved out of a very weak ``white spectrum'' source.
Each oscillator in the wave spectrum can be characterized by a small nonlinear frequency shift (cf. Fig. \ref{fig:deltaw}).
However, the wave packet that spontaneously evolves from the superposition of these oscillators sweeps
upward in frequency to maximize wave particle power exchange. While doing so, self-consistency between
chirping and rate of change of nonlinear frequency shift should be ``locked''. This is visible in Fig. \ref{fig:omura} (a),
where the intensity peak frequency (blue line) of the chorus element considered in Fig. \ref{fig:NLspectrum} 
is compared with the frequency of the corresponding peak of the rate of change 
of nonlinear frequency shift (red line). Recalling the discussion preceding Eq. (\ref{eq:WGammarenorm}), the rate of change
of the resonant frequency is $\partial_t \omega_{\rm res} = (1-v_{r\omega}/v_{g\omega})\partial_t \Delta_\omega$.
A snapshot at $\Omega_e t =2875$ of the fluctuation intensity and of the $\partial_t \omega_{\rm res}$ as a function 
of frequency is given in Fig. \ref{fig:omura} (b).
Thus, interpreting the ``whistler seeds'' of \citep{Omura2011} as the swinging oscillators in the wave packet at the intensity peak,
one should obtain the frequency increase due to the chorus chirping as 
$$ \Delta \omega = \int  (1-v_{r\omega_0(t')}/v_{g\omega_0(t')})\partial_{t'} \Delta_{\omega_0(t')} dt' \; , $$
where integration is to be intended along the red line of Fig. \ref{fig:omura} (a).
The hence obtained frequency increase is $\Delta \omega/\Omega_e = 0.24$ over the considered 
time interval, against the corresponding frequency shift, $\Delta \omega/\Omega_e = 0.21$  of the intensity peak.
Such a good agreement confirms the present explanation that reconciles the original interpretation of 
frequency chirping given by \citep{Omura2011} with the present theoretical analysis.

Further to this, and for the sake of completeness,  we would like to recall the previous discussion about the formation 
of sub-packets in connection with Fig. \ref{fig:deltaBw}. Recent statistics of 6 years of Van Allen Probes observations provided by 
\citet{Zhang2020a} have shown that the frequency variation inside sufficiently long chorus wave packets is generally finite, 
in agreement with Refs. \citep{Vomvoridis1982,Omura2008} and the present the expression, Eq. (\ref{eq:choruschirpboth}). 
However, faster frequency variations were found inside very short packets of duration less than 30 wave periods. \citet{Zhang2020a} 
explained them as due to trapping effects for relatively high amplitudes, or as due to wave superposition for  
very short packets of moderate amplitudes and duration less than 10 wave periods. 
Such statistical results have been qualitatively reproduced by numerical simulations \citep{Nunn2021}; and 
other previous works have also found some significant wave superposition 
during observations and simulations of chorus rising tones \citep{Li2011,Katoh2016,Crabtree2017}.

As a final remark, we would like to emphasize that the present theoretical analysis 
can also address some elements of the recent work by
\citep{Tsurutani2020}, based on observations using Van Allen Probe data and emphasizing that each chorus element is made
of discrete sub-elements with constant frequency. Figure \ref{fig:deltaw}, in fact, supports that each nonlinear oscillator
has a nonlinear frequency shift in the order of a few percent, consistent with observations by \citep{Tsurutani2020}.
The discrete steps, which are the essential elements of the rising tone chorus element, are instead beyond the
description of the present theoretical study since, by definition in Eq. (\ref{eq:Inorm}), we assume the continuous limit
to analytically derive the present reduced model for chorus nonlinear dynamics. Within the same theoretical framework,
it would be possible to solve the same equations in discretized form addressing, thus, the situation described by
\citep{Tsurutani2020}. This, however, is beyond the scope intended for the present work and 
hopefully will be addressed in the future. 

\section{Summary}
\label{sec:summary}

In this work, we have presented a novel and comprehensive theoretical framework of chorus wave excitation, based
on field theoretical methods introduced in \citep{Zonca2017} and in earlier works \citep{Zonca2015,Zonca2015b,Chen2016}.
This theoretical framework allows us to self-consistently
evaluate the renormalized phase space response of supra-thermal electrons, that is, the response accounting
for self-interactions in the presence of finite amplitude whistler waves. 

\begin{figure}[t]
	\centerline{\resizebox{0.9\textwidth}{!}{\includegraphics{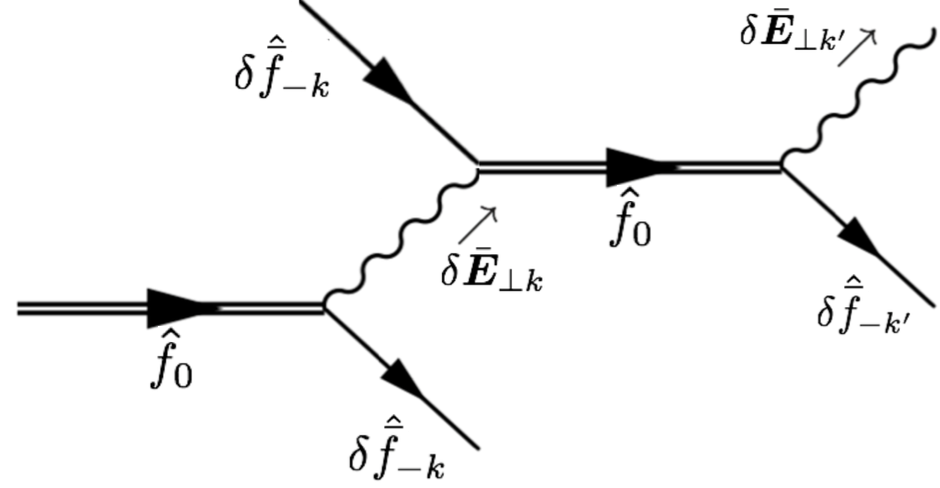}}}
\caption{Diagrammatic representation of chorus chirping consistent with the ``rules'' introduced in Fig. \ref{fig:feyndyson}.
The renormalized response of supra-thermal electrons, represented by 
the double solid line propagator, is unstable and emits and re-absorbs same-$k$ fluctuations, which is the 
strongest nonlinear process on long times. Chorus chirping occurs because, at subsequent times, different $k$'s
maximize wave particle power transfer \citep{Zonca2017}.}
\label{fig:dysonchirp}
\end{figure}

We have, furthermore, shown that the renormalized distribution function obeys a Dyson-like equation.
Since our present aim is to investigate excitation and chirping of chorus waves, we 
further simplify the Dyson-like equation by taking its velocity space moments and, ultimately, obtain equations for 
the nonlinear growth rate and frequency shifts of whistler wave packets excited by an anisotropic (bi-Maxwellian)
hot electron distribution function. Based on the structure of the hence derived governing equations, we 
analytically demonstrate for the first time that the chorus chirping rate is given by Eq. (\ref{eq:1}), originally proposed
by \citep{Vomvoridis1982}. As argued by  \citep{Vomvoridis1982,Omura2008},
chorus chirping is due to maximization of  wave particle power transfer, similar to analogous chirping 
observed in fusion plasmas \citep{Zonca2015,Zonca2015b,Chen2016}. In the light of present results,
chorus chirping can be diagrammatically illustrated as in Fig. \ref{fig:dysonchirp}.
The double solid line propagator represents the renormalized response of supra-thermal electrons,
which is unstable and, thus, nonlinearly emits oscillators belonging to the whistler spectrum. Emission 
and re-absorption of the same-$k$ has the strongest cross section \citep{Zonca2015,Zonca2015b,Chen2016}. As time
progresses, emissions are those that maximize wave particle power transfer and, thus, chirping
occurs spontaneously. 

The generality of the present theoretical approach goes well beyond the analytic derivation of \citep{Vomvoridis1982}
result of chorus chirping. It provides the insights for reconciling the present interpretation of chorus chirping with 
that originally provided by \citep{Omura2011}. It also addresses the physics underlying the evidence
of a small nonlinear frequency shift compared with the dynamic range of chorus frequency sweeping, as recently noted by \citep{Tsurutani2020}.
Meanwhile, it illuminates the origin of chorus sub-elements being the nonlinear phase modulation analogous to the
process introduced by \citep{ONeil1965}. 

The present theoretical approach also sheds light on the profound analogies of 
chorus chirping in space physics and similar non-perturbative frequency sweeping 
modes in fusion plasmas. In fact, the essential common elements are the narrow
fluctuation spectrum of chirping modes that are resonantly excited from a dense background of waves by
supra-thermal particles, which respond non-perturbatively to maximize wave particle power
transfer \citep{Zonca2015,Chen2016}.

Last but not least, this theoretical approach provides a direct proof of the one-on-one correspondence of chorus chirping with 
super-radiance in free electron lasers, noted first  by \citep{Zonca2015,Chen2016}.
It is also worthwhile emphasizing that the theoretical approaches presented in this work have interesting possible applications to 
nonlinear phenomena in high power radiation devices such as gyrotron backwave oscillators, where
they may not only be applied, but also yield in-depth understandings \citep{SChen2012,SChen2013}.

\appendix

\section{The chorus linear dispersion relation}
\label{app:linear}

Here, we briefly derive the liner dispersion relation for chorus fluctuations \citep{Kennel1966b}, emphasizing the properties
that are used for discussing the nonlinear physics addressed in this work. Reconsider Eqs. (\ref{eq:WGamma}) and (\ref{eq:powerexchange}),
and cast them as follows
\begin{eqnarray}
W(z, t, \omega)+i \Gamma(z, t, \omega) & =  & \frac{\omega_p^2}{n \omega \partial D_{w} / \partial \omega} \left\langle \frac{v_\perp^2/2}{\Omega + 
k v_\parallel - \omega} 
\left[ \frac{k}{\omega} \frac{\partial f_0}{\partial v_{ \|}}+\left(1-\frac{k v_{ \|}}{\omega}\right) \frac{1}{v_{\perp}} \frac{\partial f_0}{\partial v_{\perp}} \right] \right\rangle
\nonumber \\  & =  & \frac{\omega (\Omega - \omega)^2}{n \Omega} \left\langle \frac{v_\perp^2/2}{\Omega +  
k v_\parallel - \omega} 
\left[ \frac{k}{\omega} \frac{\partial f_0}{\partial v_{ \|}} - \frac{2}{v_\perp^2} \left(1-\frac{k v_{ \|}}{\omega}\right) f_0 \right] \right\rangle
 \; , \label{eq:WGamma4lin}
\end{eqnarray}
where, in the second line, we have integrated by parts in $v_\perp$ and used Eq. (\ref{eq:chorusdef}) to make $\partial D_{w} / \partial \omega$ explicit,
assuming $\omega^2/\omega_p^2 \ll 1$ for simplicity (cf. Sec. \ref{sec:phasespace}). Using the initial (linear) expression for $f_0$ 
given in Eq. (\ref{eq:f0eq}), the spatial dependence of density and thermal speeds are connected with the magnetic 
field non-uniformity by the condition that $f_0$ be a function of constants of motion ${\cal E} = v^2/2$ and $\mu = v_\perp^2/2B$.
The exponent in the bi-Maxwellian is then written as
$$ - \frac{\cal E}{w_\parallel^2} + \frac{\mu B_e}{w_{\perp e}^2} \frac{B}{B_e} \left( \frac{w_{\perp e}^2}{w_\parallel^2} - \frac{w_{\perp e}^2}{w_\perp^2} \right) \; .$$
For this to be constant for arbitrary ${\cal E} = v^2/2$ and $\mu = v_\perp^2/2B$, we need to set $w_\parallel = w_{\parallel e}$ and $w_\perp = \zeta w_{\perp e}$,
with $\zeta^{-2} =  1 + A (1 - B_e/B) = 1 + A \xi z^2/(1 + \xi z^2)$, as noted below  Eq. (\ref{eq:f0eq}). Furthermore, the pre-factor in the
 bi-Maxwellian is constant only if $n_0/n_e = w_\perp^2/ w_{\perp e}^2 = \zeta^2$.  Meanwhile,
performing the velocity space integration, it is possible to write
\begin{eqnarray}
W(z, 0, \omega)+i \Gamma(z, 0, \omega) & = & \frac{n_e(\Omega - \omega)^2}{n \Omega} \zeta^2 \left[ 1 - (A+1) \zeta^2 + (A+1) \zeta^2 \frac{(\Omega - \omega)}{\sqrt{2}|k| w_{\parallel e}} Z\left( \frac{\omega-\Omega}{\sqrt{2}|k| w_{\parallel e}} \right) \right. \nonumber \\
& & \left.- \frac{\Omega}{\sqrt{2}|k| w_{\parallel e}} 
Z\left( \frac{\omega-\Omega}{\sqrt{2}|k| w_{\parallel e}} \right) \right] \; . \label{eq:WGammalin0}
\end{eqnarray}
Here, from Sec. \ref{sec:waveequations}, we have recalled  $n_0 = \zeta^2 n_e$ and $Z(x) = \pi^{-1/2} \int_{-\infty}^\infty e^{-y^2}/(y-x) dy$ is the plasma
dispersion function. Noting $A+1 - \zeta^{-2} = A/(1+\xi z^2)$, Eq. (\ref{eq:WGammalin0}) can be rewritten as
\begin{equation}
W(z, 0, \omega)+i \Gamma(z, 0, \omega) = \frac{n_e(\Omega - \omega)^2}{n \Omega} \zeta^4 \left[ - \frac{A}{1+\xi z^2} +  \frac{A\Omega_e - (A+1)\omega}{\sqrt{2}|k| w_{\parallel e}} Z\left( \frac{\omega-\Omega}{\sqrt{2}|k| w_{\parallel e}} \right) \right] \; . \label{eq:WGammalin}
\end{equation}
This expression shows that $\omega/\Omega_e = A/(A+1)$ is the frequency where wave particle power exchange with hot electrons
changes sign and the driving rate becomes a damping \citep{Kennel1966b}. Meanwhile, Eq. (\ref{eq:WGammalin}) also shows
that $W(z, 0, \omega)$ and $\Gamma(z, 0, \omega)$ scale as $\zeta^4$. Thus, recalling 
from Sec. \ref{sec:waveequations} that $\zeta^{-2} = 1 + A \xi z^2/(1+\xi z^2)$, the length scale of the hot electron contribution
to the chorus dispersion relation is $\sim (A\xi)^{-1/2}$, which, already at moderate values of 
$A$, rapidly takes over the non-uniformity due to the ambient magnetic field,
that is $B = B_e (1 + \xi z^2)$ \citep{Helliwell1967}, as illustrated in Fig. \ref{fig:WGamma}.
This suggests that formal simplification can be achieved in the analytical investigation of chorus nonlinear 
dynamics, addressed in this work, by assuming a non-uniform source of hot electrons, localized about
the equator, neglecting, meanwhile, magnetic field non-uniformity. 
As noted in Sec. \ref{sec:waveequations}, this assumption, although not strictly necessary, helps simplifying 
the analytical derivations in this work; and can be formally obtained for $A\gg 1$. In fact, as noted above, the
$\sim (A\xi)^{-1/2}$ length scale in the hot electron non-uniform response takes over the magnetic field non-uniformity already
at moderate values of $A$.

\section{The Dyson-Schwinger equation approach}
\label{app:dyson}

Here, we elaborate the Dyson-Schwinger equation approach presented in Refs. \citep{Chen2016,Zonca2015},
with the applications to magnetized fusion plasma presented therein, specializing it to 
nonlinear dynamics and phase space transport by chorus emission. 

\begin{figure}[t]
	\centerline{\resizebox{\textwidth}{!}{\includegraphics{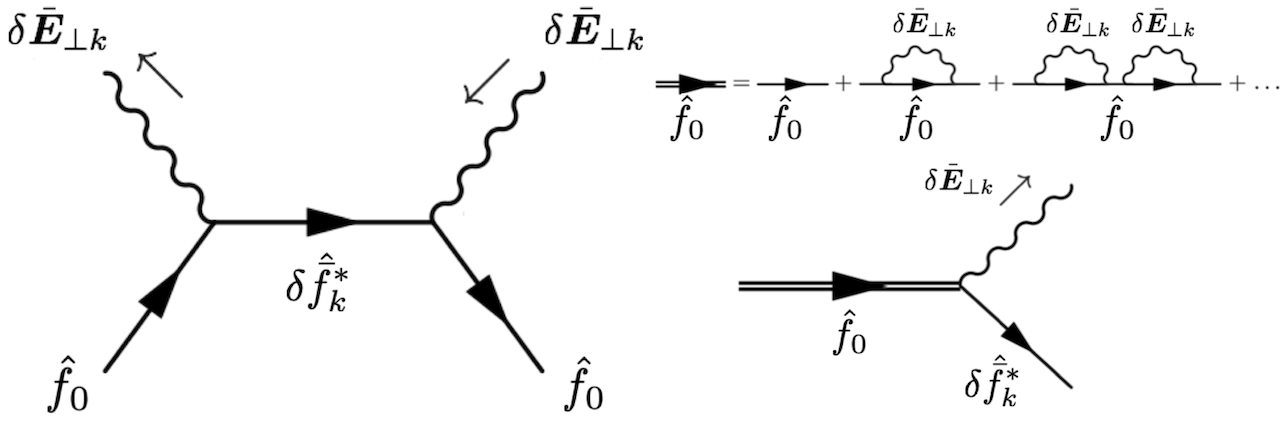}}} \
\vspace*{-2em}\newline\noindent(a) \hspace*{0.48\linewidth} (b)
\caption{(a) Diagrammatic representation of the elementary processes of Eqs. (\ref{eq:dbarfkvla1}) and (\ref{eq:barf0vla1}). (b) Diagrammatic representation of the 
renormalized $\hat f_0$ hot electron response as solution of the Dyson-like equation (\ref{eq:barf0vla1}).}
\label{fig:feyndyson}
\end{figure}

\begin{sloppypar} The ``Dyson-like equation'' terminology, by analogy
with the earlier work by \citep{Altshul1966}, was introduced by \citep{Chen2016,Zonca2015} as a tribute
to Freeman J. Dyson, who recently passed away (https://en.wikipedia.org/wiki/Freeman\_Dyson). 
The Dyson-Schwinger equations,
as equations of motion of Green functions, provide a complete description of the theory \citep{Dyson1949}, since they describe
the propagation as well as interaction of the fields themselves. From this point of view, Dyson-Schwinger equations,
and Eq. (\ref{eq:barf0vla1}) as a particular case, can be used to generate perturbation expansions 
in the weak field limit, cf. Fig. \ref{fig:feyndyson} (b),
but can also be adopted for the more general strong-coupling case. \end{sloppypar}
The elementary process that underlies this dynamics is illustrated in Fig. \ref{fig:feyndyson} (a), 
where we have borrowed and suitably modified the Feynman diagram rules as in \citep{Chen2016, Zonca2017} 
to illustrate Eq. (\ref{eq:dbarfkvla1}) and its reverse.
In particular, straight lines represent
linearized propagators (Green functions) of particle distribution functions, while wavy lines stand for linearized propagators (Green functions)
of fluctuating electromagnetic fields.
Arrows indicate the direction of propagation. Meanwhile, nodes represent (nonlinear) interactions/couplings. 
Furthermore, because of energy and momentum conservation in particle and electromagnetic fields 
field interactions, propagation of fields is equivalent to the opposite propagation of 
corresponding complex conjugate fields \citep{Zonca2015b}. For example, emission of $\delta \bar E_k$ corresponds to 
absorption of  $\delta \bar E_k^*$ because of symmetry under parity and time reversal transformations. Thus,  the left node (vertex) in Fig. \ref{fig:feyndyson} (a)
represents (the $c.c.$ of) Eq. (\ref{eq:dbarfkvla1}); while the right node (vertex) represents the first two lines of Eq. (\ref{eq:barf0vla1}).
In the present theoretical approach, emission and reabsorption of $\delta \bar E_k$ and $\delta \bar E_k^*$ 
can occur repeatedly. Here, by emission we mean ``generation of waves'' because of the instability driven by the spatially averaged
electron distribution function $\hat f_0$. Meanwhile, by reabsorption we intend to mean the ``nonlinear interaction'' of electromagnetic
fluctuations with the perturbed electron distribution function that modifies $\hat f_0$ itself. 
This is illustrated
in the upper part of Fig. \ref{fig:feyndyson} (b) in the form of a Dyson series, and dominates the nonlinear dynamics since it can be shown to cause the most significant
distortion of $\hat f_0$ on the long time scale \citep{Vanhove1954,Prigogine1962,Balescu1963,Altshul1966,Dupree1966,
Aamodt1967,Weinstock1969,Mima1973,Zonca2015, Zonca2015b, Chen2016, Zonca2017}. 
Such a distortion of the hot electron distribution function, determined self-consistently in the presence of the finite amplitude
fluctuation spectrum, constitutes the ``renormalized'' hot electron response, denoted with the double solid line 
in Fig. \ref{fig:feyndyson} (b). It is this renormalized hot electron $\hat f_0$, which is evolving in time, that 
self-consistently causes the evolution of the fluctuation spectrum according to Eqs. (\ref{eq:actionevolve}) to (\ref{eq:WGamma}) and as illustrated 
in the lower part of Fig. \ref{fig:feyndyson} (b).
Equation (\ref{eq:barf0vla1}), meanwhile, is a nonlinear integro-differential equation and can be used to close the chorus wave equations discussed in Sec. \ref{sec:waveequations}. In fact, it describes the response of the $k=0$ hot electron distribution function 
by continuous emission and reabsorption of whistler waves,
shown in Fig. \ref{fig:feyndyson} (b), which are amplified due to wave particle resonant interactions. Again, we 
note that this emission and reabsorption occur with
any generic whistler wave packet as denoted by the summation over the whole fluctuation spectrum, which is evolving in time  self-consistently with the
$k=0$ particle distribution function.
In this respect, as noted already, Eq. (\ref{eq:barf0vla1}) can be viewed as the renormalized hot electron distribution function evolving 
on the nonlinear time scale, which justifies dubbing it as Dyson-like equation \citep{Chen2016, Zonca2015, Zonca2015b}. 

\section{Detailed derivation of Eqs. (\ref{eq:redWGammabar0}) and (\ref{eq:redWbar0}).} 

\subsection{Derivation of Eq. (\ref{eq:redWGammabar0}).}
\label{app:nonlinear}

In this Appendix, we briefly summarize the derivation of Eq. (\ref{eq:redWGammabar0}) from Eq. (\ref{eq:redWGammabar}) based on 
Eqs. (\ref{eq:barf0vla1}) and (\ref{eq:forinv1}). For consistency with Sec. \ref{sec:reddyson}, we also denote the current frequency 
and wave number satisfying the lowest order whistler wave dispersion relation 
as $\bar \omega$ and $\bar k$ in order to distinguish them from $\omega$ and $k$ 
in the running summation over the fluctuation spectrum. Formally, we can rewrite Eq. (\ref{eq:redWGammabar}) as
\begin{eqnarray}
\bar W (\bar \omega) +i \bar \Gamma (\bar \omega) & =  & \frac{n_e}{n} \left( 1 - \frac{v_{r\bar\omega}}{v_{g\bar\omega}} \right)^2
\left\langle \frac{v_\perp^2}{2} \Omega_e \left[ \Omega_e + \bar k v_\parallel - \bar \omega - i (1-v_{r\bar\omega}/v_{g\bar\omega}) \partial_t\right]^{-1}   \right. \nonumber \\
& & \times \left. \left( \frac{\bar k}{\bar \omega} \frac{\partial}{\partial v_\parallel} - \frac{2}{v_\perp^2} \frac{\Omega_e}{\bar \omega} \right)  \left( \partial_t + v_\parallel \partial_z \right)^{-1}  \left( \partial_t + v_\parallel \partial_z \right) \hat f_0 \right\rangle \; , \label{eq:redWGammabar0app0}
\end{eqnarray}
where, for brevity, we have omitted the dependences on $t-z/v_{g\bar\omega}$.
From this, upon substitution of Eq. (\ref{eq:barf0vla1}) and noting Eq. (\ref{eq:forinv1}), we have
\begin{eqnarray}
\bar W (\bar \omega) +i \bar \Gamma (\bar \omega) & =  & \frac{n_e}{n} \left( 1 - \frac{v_{r\bar\omega}}{v_{g\bar\omega}} \right)^2
\left\langle \frac{v_\perp^2}{2} \Omega_e \left[ \Omega_e + \bar k v_\parallel - \bar \omega - i (1-v_{r\bar\omega}/v_{g\bar\omega}) \partial_t\right]^{-1}   \right. \nonumber \\
& & \times  \left( \frac{\bar k}{\bar \omega} \frac{\partial}{\partial v_\parallel} - \frac{2}{v_\perp^2} \frac{\Omega_e}{\bar \omega} \right)  \sum_k  \frac{e^2}{2 m^2}  v_\perp |\delta \bar E_k| \left[ \frac{k}{\omega}\frac{\partial}{\partial v_\parallel}
+ \left( 1 - \frac{k v_\parallel}{\omega} \right) \left( \frac{1}{v_\perp} \frac{\partial}{\partial v_\perp} + \frac{1}{v_\perp^2} \right) \right] \nonumber \\
& & \times  \left[ (\Omega_e + k v_\parallel - \omega)^2 + (1-v_{r\omega}/v_{g\omega})^2 \partial_t^2\right]^{-1} (1-v_{r\omega}/v_{g\omega}) \partial_t  v_\perp |\delta \bar E_k| \nonumber \\ & & \times \left.  
\left[ \frac{k}{\omega}\frac{\partial}{\partial v_\parallel}
+ \left( 1 - \frac{k v_\parallel}{\omega} \right) \frac{1}{v_\perp} \frac{\partial}{\partial v_\perp} \right]  \hat f_0 \right\rangle \nonumber \\
& \simeq & \frac{n_e}{n} \left( 1 - \frac{v_{r\bar\omega}}{v_{g\bar\omega}} \right)^2
\left\langle \frac{v_\perp^2}{2} \Omega_e \left[ \Omega_e + \bar k v_\parallel - \bar \omega - i (1-v_{r\bar\omega}/v_{g\bar\omega}) \partial_t\right]^{-1}   \right. \nonumber \\
& & \times  \left( \frac{\bar k}{\bar \omega} \frac{\partial}{\partial v_\parallel} - \frac{2}{v_\perp^2} \frac{\Omega_e}{\bar \omega} \right)  \sum_k  \frac{e^2}{2 m^2}  v_\perp |\delta \bar E_k| \left[ \frac{k}{\omega}\frac{\partial}{\partial v_\parallel}
+ \frac{\Omega_e}{\omega} \left( \frac{1}{v_\perp} \frac{\partial}{\partial v_\perp} + \frac{1}{v_\perp^2} \right) \right] \nonumber \\
& & \times  \left[ (\Omega_e + k v_\parallel - \omega)^2 + (1-v_{r\omega}/v_{g\omega})^2 \partial_t^2\right]^{-1} (1-v_{r\omega}/v_{g\omega}) \partial_t  v_\perp |\delta \bar E_k| \nonumber \\ & & \times \left.  
\left[ \frac{k}{\omega}\frac{\partial}{\partial v_\parallel}
+ \frac{\Omega_e}{\omega} \frac{1}{v_\perp} \frac{\partial}{\partial v_\perp} \right]  \hat f_0 \right\rangle
 \;  . \label{eq:redWGammabar0app1}
\end{eqnarray}
To derive Eq. (\ref{eq:redWGammabar0}), the last step is to integrate by parts twice in $v_\perp$ in order to eliminate $\partial_{v_\perp}$, taking into account that 
\begin{equation}
\left\langle v_\perp^4 \hat f_0 \right\rangle = 2 \frac{\left\langle v_\perp^2 \hat f_0 \right\rangle^2}{\left\langle \hat f_0 \right\rangle} \; , \label{eq:perpmaxw}
\end{equation}
for the anisotropic Maxwellian of Eq. (\ref{eq:f0eq}).

\subsection{Derivation of Eq. (\ref{eq:redWbar0}).}
\label{app:nonlinear2}

 Velocity space integration in Eq. (\ref{eq:redWGammabar0}) is naturally (and more rigorously) performed in the Laplace- rather than in the time-representation  
\citep{Chen2016, Zonca2015, Zonca2017, Tao2020}. Here, however, for the sake of simplicity and conciseness, we directly 
manipulate Eq. (\ref{eq:redWGammabar0}) in the time-representation formally handling operator symbols. 
Let's first note, considering Eq. (\ref{eq:forinv0}),
\begin{eqnarray}
\left[ \Omega_e + \bar k v_\parallel - \bar \omega - i (1-v_{r\bar\omega}/v_{g\bar\omega}) \partial_t\right]^{-1} 
 & \simeq & \left[ (\Omega_e + \bar k v_\parallel - \bar \omega)^2 
+  (1-v_{r\bar\omega}/v_{g\bar\omega})^2 \partial_t^2 \right]^{-1} \nonumber \\
& & \times \left[ \Omega_e + \bar k v_\parallel - \bar \omega + i (1-v_{r\bar\omega}/v_{g\bar\omega}) \partial_t \right]
\nonumber \\
 & \simeq & \left[ x^2 + a^2 \right]^{-1} \left[ x + i a\right] \; , \label{eq:forinv2}
\end{eqnarray}
having denoted symbolically $x \equiv \bar k (v_\parallel - v_{r\bar\omega})$ and $a \equiv (1-v_{r\bar\omega}/v_{g\bar\omega}) \partial_t$.
Here, $a$ as an operator is meant to be acting on $|\delta \bar E_{\bar k}|$ and what follows in the representation of the 
integrand, 
that is $|\delta \bar E_{k}|^2$ and $\hat f_0$. In fact, in the derivation
of Eqs. (\ref{eq:powerexchange}) and (\ref{eq:redWGamma}), we have normalized the wave particle power exchange to 
$|\delta \bar E_{\bar k}|^2$. Thus,  
the nonlinear frequency and wave 
number shift due to the incremental change in the wave packet amplitude
and phase is reabsorbed into $\bar k (v_\parallel - v_{r\bar\omega})$ as will be further discussed below in Sec. \ref{sec:choruschirp}.
We can adopt the same symbolic representation for Eq. (\ref{eq:forinv1}) and, thus,
\begin{equation}
\left[ (\Omega_e + k v_\parallel - \omega)^2 + (1-v_{r\omega}/v_{g\omega})^2 \partial_t^2\right]^{-1}  (1-v_{r\omega}/v_{g\omega}) \partial_t 
\simeq \left[ (x-x_0)^2 + b^2 \right]^{-1} b  \; , \label{eq:forinv3}
\end{equation}
where $x_0 \equiv \omega - \bar \omega - v_{r\omega} (k - \bar k) \simeq (1-v_{r\omega}/v_{g\omega}) (\omega - \bar \omega)$ and $b \equiv (1-v_{r\omega}/v_{g\omega}) \partial_t$ as an operator is meant to be acting on $|\delta \bar E_{ k}|$
and $\hat f_0$. Meanwhile, for resonant particles, we note that, for $\mathbb{R} e a>0, \mathbb{R} e b>0 \text { and }|a|,|b|,|x_0| \ll 1$
\begin{eqnarray}
\left[ x^2 + a^2 \right]^{-1} a  \left[ (x-x_0)^2 + b^2 \right]^{-1} b & \simeq & (\pi / 2) \left[x_0^2 + (a+b)^2\right]^{-1} (a+b) \left( \delta(x) + \delta (x-x_0) \right) 
\label{eq:urel1} \\ & &  - (\pi / 2) \left[x_0^2 + (a-b)^2\right]^{-1} (a-b) \left( \delta(x) - \delta (x-x_0) \right) \; ,
\nonumber
\end{eqnarray}
inside the velocity space integrand; and
\begin{eqnarray}
\left[ x^2 + a^2 \right]^{-1} x \left[ (x-x_0)^2 + b^2 \right]^{-1} b & \simeq & (\pi / 2) \left[x_0^2 + (a+b)^2\right]^{-1} x_0 \left( \delta(x) + \delta (x-x_0) \right) 
\label{eq:urel2} \\ & &  - (\pi / 2) \left[x_0^2 + (a-b)^2\right]^{-1} x_0 \left( \delta(x) - \delta (x-x_0) \right) \; . \nonumber
\end{eqnarray}
Terms depending on $\delta(x)$ in the velocity space integrand are computed at $v_\parallel = v_{r\omega}$, while those depending on 
$\delta(x-x_0)$ are computed at $v_\parallel = v_{r\bar\omega}$.
We can further simplify these expressions noting that, for a narrow spectrum, contributions $\propto \left( \delta(x) - \delta (x-x_0) \right)$ 
can be neglected. Furthermore, the symbolic expression of $(a + b)$ can also be simplified noting assumption (i) above and, thus
\begin{eqnarray}
(a+b) & \simeq & |\delta \bar E_{ k}|^{-3} \hat f_0^{-1} (1-v_{r\omega}/v_{g\omega}) \partial_t |\delta \bar E_{ k}|^3 \hat f_0  \nonumber \\
& & + 
 |\delta \bar E_{ k}|^{-1} \hat f_0^{-1} (1-v_{r\omega}/v_{g\omega}) \partial_t |\delta \bar E_{ k}| \hat f_0 \nonumber \\
 & \simeq &  2 |\delta \bar E_{ k}|^{-2} \hat f_0^{-1}  (1-v_{r\omega}/v_{g\omega}) \partial_t |\delta \bar E_{ k}|^2 \hat f_0 \; . \label{eq:urel3}
 \end{eqnarray}
Again, these symbolic relations are more rigorously interpreted in the Laplace- rather than the time-representation \citep{Chen2016, Zonca2015, Zonca2017, Tao2020}.
Interested readers are referred to the original references for more details.

Based on these relations, one can derive Eqs. (\ref{eq:redWbar0}) and (\ref{eq:redGammabar0}), where 
we have noted that
\begin{equation}
\frac{1}{k} \frac{\partial}{\partial v_{r\omega}} = \frac{1}{(1-v_{r\omega}/v_{g\omega})} \frac{\partial}{\partial \omega} \; ,
\label{eq:urel4}
\end{equation}
justifying the usefulness of introducing the normalization of $\bar W$ and $\bar \Gamma$ as in Eq. (\ref{eq:redWGammabar}) in Sec. \ref{sec:phasespace}. The presence of the integral operator
\begin{equation}
\left[ \frac{\left( \omega - \bar \omega \right)^2}{4 \Omega_e^2}+\Omega_e^{-2} \partial_t^2\right]^{-1} \label{eq:integrop}
\end{equation}
is what allows us to neglect the non-resonant particle response (Cauchy principal value) in the derivations above.

\section{Evolution equations for numerical solution of the reduced Dyson-like equation.} 
\label{app:numsol}

To exploit the dense nature of the whistler wave spectrum, we introduce the dimensionless intensity ${\cal I} (\omega)$ such that
\begin{equation}
\sum_k \frac{\left\langle\left\langle \omega_{{\rm tr}k}^4 \right\rangle\right\rangle}{\Omega_e^4} = \frac{\hat \gamma_e^2}{\Omega_e^2} \int  \frac{d \omega}{\Omega_e} {\cal I} (\omega) \frac{\omega^{3/2}}{(\Omega_e - \omega)^{3/2}} \; . \label{eq:Inorm}
\end{equation}
Here, the $\omega^{3/2}/(\Omega_e - \omega)^{3/2}$ factor accounts for the 
$\sim I_k k^3$ scaling of $\left\langle\left\langle \omega_{{\rm tr}k}^4 \right\rangle\right\rangle$. This normalization is chosen ad hoc to have nonlinearity effects being important when ${\cal I} (\omega) \sim {\cal O}(1)$. 

\begin{figure}[t]
	\centerline{\resizebox{\textwidth}{!}{\includegraphics{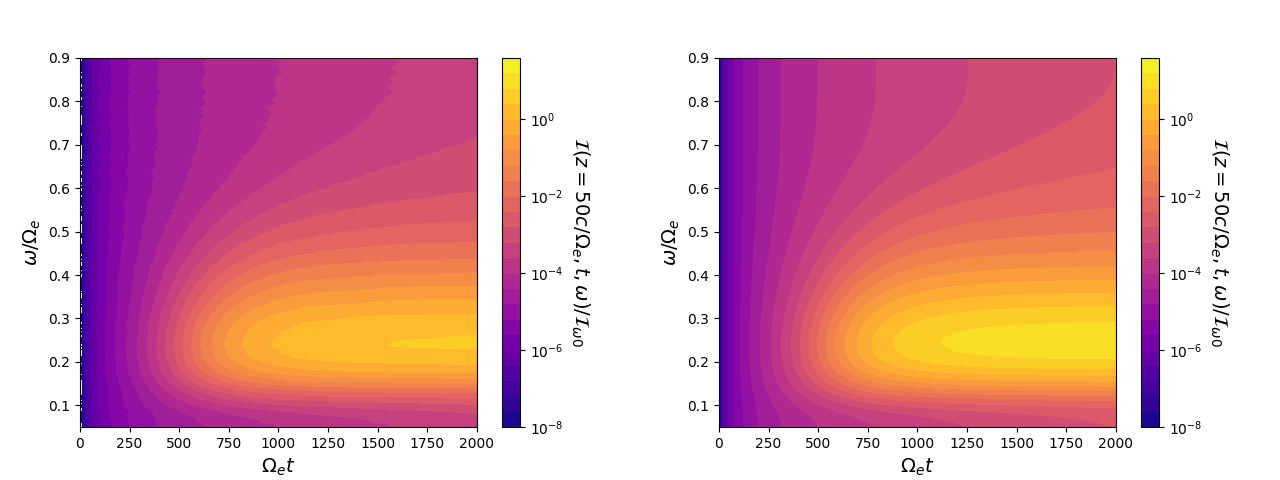}}} \
\vspace*{-2em}\newline\noindent(a) \hspace*{0.48\linewidth} (b)
\caption{Contour plot of the linear evolution of ${\cal I} (z=50 c/\omega_e,t,\omega)$ for (a) a random source and (b) a uniform source of the same strength ${\cal S} = 2 \times 10^{-5}$  in Eq. (\ref{eq:ievolve}). The position is fixed at $z \Omega_e/c = 50$ and parameters are the same as in Fig. \ref{fig:WGamma}.}
\label{fig:rancoh}
\end{figure}

To invert the integral operator of Eq. (\ref{eq:integrop}), let us introduce the auxiliary functions
\begin{eqnarray}
& &\left[ \frac{\left( \omega - \bar \omega \right)^2}{4 \Omega_e^2}+\Omega_e^{-2} \partial_t^2\right] G_{L1} (\omega,\bar \omega) 
= \frac{\hat \gamma_e^2}{\Omega_e^2} \frac{{\cal I} (\omega)}{(1-v_{r\omega}/v_{g\omega})^4} \bar \Gamma_L (\omega) \nonumber \; , \\
& &\left[ \frac{\left( \omega - \bar \omega \right)^2}{4 \Omega_e^2}+\Omega_e^{-2} \partial_t^2\right] G_{L2} (\omega,\bar \omega) 
= \frac{\hat \gamma_e^2}{\Omega_e^2} \frac{{\cal I} (\omega)}{(1-v_{r\omega}/v_{g\omega})^4} \bar \Gamma_L (\bar \omega) \nonumber \; , \\
& &\left[ \frac{\left( \omega - \bar \omega \right)^2}{4 \Omega_e^2}+\Omega_e^{-2} \partial_t^2\right] G_{NL1} (\omega,\bar \omega) 
= \frac{\hat \gamma_e^2}{\Omega_e^2} \frac{{\cal I} (\omega)}{(1-v_{r\omega}/v_{g\omega})^4} \bar \Gamma_{NL} (\omega) \nonumber \; , \\
& &\left[ \frac{\left( \omega - \bar \omega \right)^2}{4 \Omega_e^2}+\Omega_e^{-2} \partial_t^2\right] G_{NL2} (\omega,\bar \omega) 
= \frac{\hat \gamma_e^2}{\Omega_e^2} \frac{{\cal I} (\omega)}{(1-v_{r\omega}/v_{g\omega})^4} \bar \Gamma_{NL} (\bar \omega) \label{eq:Gdefs} \; ;
\end{eqnarray}
where, from Eq. (\ref{eq:redGammabar0}),
\begin{eqnarray}
\bar \Gamma_{NL} (\bar \omega) & =  & \left[ \Omega_e \frac{\partial}{\partial \bar \omega} - \frac{2 \Omega_e^2}{\bar k^2 \left\langle v_\perp^2\right\rangle} 
\left( 1 - \frac{v_{r\bar\omega}}{v_{g\bar\omega}} \right) \right] \Omega_e \frac{\partial}{\partial \bar \omega}  \int  \frac{d \omega}{\Omega_e} \frac{\omega^{3/2}}{(\Omega_e - \omega)^{3/2}} \nonumber \\
& & \times  \frac{1}{8} \left[ G_{L1} (\omega,\bar \omega) + G_{L2} (\omega,\bar \omega)
+ G_{NL1} (\omega,\bar \omega) + G_{NL2} (\omega,\bar \omega) \right]\label{eq:redGammabar2} \; .
\end{eqnarray}
Meanwhile, Eq. (\ref{eq:redWbar0}) can be cast as
\begin{eqnarray}
\Omega_e^{-1} \partial_t \bar W (\bar \omega) & =  & \left[ \Omega_e \frac{\partial}{\partial \bar \omega} - \frac{2 \Omega_e^2}{\bar k^2 \left\langle v_\perp^2\right\rangle} 
\left( 1 - \frac{v_{r\bar\omega}}{v_{g\bar\omega}} \right) \right] \Omega_e \frac{\partial}{\partial \bar \omega}  \int  \frac{d \omega}{\Omega_e} \frac{\omega^{3/2}}{(\Omega_e - \omega)^{3/2}} \frac{\left( \omega - \bar \omega \right)}{2 \Omega_e} \nonumber \\
& & \times  \frac{1}{8} \left[ G_{L1} (\omega,\bar \omega) + G_{L2} (\omega,\bar \omega)
+ G_{NL1} (\omega,\bar \omega) + G_{NL2} (\omega,\bar \omega) \right]\label{eq:redWbar1} \; .
\end{eqnarray}
Equations (\ref{eq:Gdefs}) to (\ref{eq:redWbar1}) are closed by the intensity evolution equation,
\begin{eqnarray}
\Omega_e^{-1} \partial_t {\cal I} (\omega) & = & 2 {\cal S} {\cal I} (\omega)^{1/2} + 2 {\cal I} (\omega) \frac{\omega(\Omega_e-\omega)^2}{\Omega_e^3} \left[ \partial_t  \left( \int_{z-v_{g\omega}t}^\infty \zeta^4 (z') \frac{dz'}{v_{g\omega}} \right) \bar \Gamma(\omega) \right. \nonumber \\
& &\left.  \left.+ \left( \int_{z-v_{g\omega}t}^\infty \zeta^4 (z') \frac{dz'}{v_{g\omega}} \right) \partial_t \bar \Gamma (\omega) \right]\right/ \left( 1 - \frac{v_{r\omega}}{v_{g\omega}} \right)^2\; ; \label{eq:ievolve}
\end{eqnarray}
and the wave packet phase evolution equation, 
\begin{eqnarray}
\Omega_e^{-1} \partial_t \varphi (\omega) & = & - \frac{\omega(\Omega_e-\omega)^2}{\Omega_e^3} \left[ \partial_t  \left( \int_{z-v_{g\omega}t}^\infty \zeta^4 (z') \frac{dz'}{v_{g\omega}} \right) \bar W (\omega) \right. \nonumber \\
& &\left. \left.  + \left( \int_{z-v_{g\omega}t}^\infty \zeta^4 (z') \frac{dz'}{v_{g\omega}} \right) \partial_t \bar W (\omega) \right]\right/ \left( 1 - \frac{v_{r\omega}}{v_{g\omega}} \right)^2 \; ; \label{eq:phievolve}
\end{eqnarray}
which can be readily derived from Eqs. (\ref{eq:actionevolve}) and (\ref{eq:phaseevolve}) keeping in mind the discussion 
given in the first paragraph of Sec. \ref{sec:reddyson}.
Note that we have added a source term $2 {\cal S} {\cal I} (\omega)^{1/2}$ on the right hand side of Eq. (\ref{eq:ievolve}). 
The value of ${\cal S}$ represents the injection rate 
of fluctuations in the $|\delta \bar E_k|$ spectrum. We adopted it because it gives us a variety of possibilities rather than assuming an initial spectrum; \exgra, using ${\cal S}$ as a random source stirring the system or a constant uniform source. Figure \ref{fig:rancoh} gives a comparison of the linear evolution of ${\cal I} (z=50 c/\omega_e,t,\omega)$ in the case of a random (a) and uniform source (b) of the same strength ${\cal S} = 2 \times 10^{-5}$. Parameters are the same as in Fig. \ref{fig:WGamma}. In both cases, we clearly note the predominance of a narrow spectrum at the most unstable frequency after the whistler wave packet has been convectively amplified by crossing the  localized hot electron source at the equator. Because of this, we will focus on the uniform source case in the following, and discuss random or more general sources in later studies.
Equations (\ref{eq:Gdefs}) to (\ref{eq:phievolve}) fully characterize the self-consistent nonlinear evolution of a whistler wave packet
spectrum excited by wave particle resonances with a hot electron source that is  localized about the equator. As anticipated in
Secs. \ref{sec:phasespace} and \ref{sec:reddyson}, they assume that the whistler wave packets belong to a dense
(nearly continuous) spectrum that are continuously emitted and reabsorbed (cf. Fig. \ref{fig:feyndyson} and 
corresponding discussion) to maximize wave particle power transfer 
\citep{Chen2016, Zonca2015}. When solving these equations by advancing them in time by
one step $\Delta t$, one has to keep in mind that the wave packet amplitude and phase are shifted nonlinearly and
cause  hot electron induced frequency ($\Delta(\Delta_\omega) = - \Delta t \partial_t^2 \varphi(\omega)$) and 
corresponding wave number shifts, introduced in Sec. \ref{sec:phasespace}. Despite $\Delta_\omega/\Delta_k = v_{g\omega}$,
and, thus, the shifted wave packet still satisfies the whistler dispersion relation, the hot electron induced phase shift causes 
a corresponding small but finite shift in the resonant velocity $k \Delta v_{r\omega} = \Delta_\omega - v_{r\omega} 
\Delta_k = (1-v_{r\omega}/v_{g\omega})\Delta_\omega$. This means that, after advancing
in time by one step $\Delta t$, the functions $\bar \Gamma_{NL}$ and $\bar W_{NL}$ are actually evaluated at
$\bar \omega + \Delta(\Delta_\omega)$. Thus, noting Eq. (\ref{eq:urel4}), these functions have to be updated as 
\begin{eqnarray}
\bar \Gamma_{NL} (\bar \omega) & \rightarrow & \bar \Gamma_{NL} (\bar \omega) - 
\Delta(\Delta_\omega) \frac{\partial}{\partial \bar \omega} \bar \Gamma_{NL} (\bar \omega) \; , \nonumber \\
\bar W_{NL} (\bar \omega) & \rightarrow & \bar W_{NL} (\bar \omega) - 
\Delta(\Delta_\omega) \frac{\partial}{\partial \bar \omega} \bar W_{NL} (\bar \omega) \; . \label{eq:WGammarenorm}
\end{eqnarray}
A similar argument applies to ${\cal I}(\omega)$ and $\varphi(\omega)$, such that
\begin{eqnarray}
{\cal I} (\omega) & \rightarrow & {\cal I} (\omega) - 
\Delta(\Delta_\omega) \frac{\partial}{\partial \omega} {\cal I} (\omega) \; , \nonumber \\
\varphi (\omega) & \rightarrow & \varphi (\omega) - 
\Delta(\Delta_\omega) \frac{\partial}{\partial \omega} \varphi (\omega) \; ; \label{eq:Iphirenorm}
\end{eqnarray}
after each time step. Note that Eq. (\ref{eq:Iphirenorm}) corresponds to changing $\partial_t \rightarrow \partial_t - \partial_t^2 \varphi(\omega) \partial_\omega$
on the left hand side of Eqs. (\ref{eq:ievolve}) and (\ref{eq:phievolve}); 
that is, to solving the wave kinetic equation \citep{Bernstein1977,McDonald1988} for wave packet intensity and phase.

Considering ${\cal I} (\omega) = 0$ together with  $\varphi (\omega) = 0$ at $t=0$, initial conditions for Eqs. (\ref{eq:Gdefs}) are particularly simple: 
\begin{eqnarray}
\left. G_{...}(\omega, \bar \omega) \right|_{t=0} & =  & 0 \; , \nonumber \\ 
\left. \partial_t G_{...}(\omega, \bar \omega) \right|_{t=0} & =  & 0 \; ; \label{eq:initG}
\end{eqnarray}
where, for brevity, we have generically denoted by $\ldots$ the subscripts of $G_{...}(\omega, \bar \omega)$ 
functions defined in Eqs. (\ref{eq:Gdefs}). Furthermore, for Eqs. (\ref{eq:redGammabar2}) and (\ref{eq:redWbar1}), 
\begin{eqnarray}
\left. \bar \Gamma_{NL}(\omega) \right|_{t=0} & =  & 0 \; , \nonumber \\ 
\left.  \bar W_{NL}(\omega) \right|_{t=0} & =  & 0 \; . \label{eq:initWGbar}
\end{eqnarray}
Finally, considering a frequency domain such that the fluctuation spectrum at the boundary is sufficiently small 
that it can be neglected, boundary conditions are the trivial ones; i.e., $G_{...}(\omega, \bar \omega)$ functions, $\bar \Gamma_{NL}(\omega)$ and
$\bar W_{NL}(\omega)$ vanish at any time. However, more realistically and to avoid undesirable discontinuities 
at the boundary of the frequency simulation domain, we assume that values at the boundaries are obtained by linear
extrapolation of the inner solution.

\section*{Acknowledgments}
This work was carried out within the framework of the EUROfusion Consortium 
and received funding from Euratom research and training programme 2014--2018 
and 2019--2020 under Grant Agreement No. 633053 (Project No. WP19-ER/ENEA-05). 
The views and opinions expressed herein do not necessarily reflect those of the European Commission.
This work was also supported by NSFC grants (41631071 and 11235009), the Strategic Priority Program 
of the Chinese Academy of Sciences (No. XDB41000000)
and the Fundamental Research Funds for the Central Universities. 
Code and input files used for generating the data used in this study can be found at
https://doi.org/10.5281/zenodo.5076015.


%
%

\bibliography{xt}

%
%
%
%
%

\end{document}


%
%


\title{Supporting Information for "Insert Title"}
%
%

%
%



\authors{=Authors=}


\affiliation{=number=}{=Affiliation Address=}

%
%

%

\begin{article}

%
%

\noindent\textbf{Contents of this file}
\begin{enumerate}
\item Text S1 to Sx
\item Figures S1 to Sx
\item Tables S1 to Sx
\end{enumerate}
\noindent\textbf{Additional Supporting Information (Files uploaded separately)}
\begin{enumerate}
\item Captions for Datasets S1 to Sx
\item Captions for large Tables S1 to Sx (if larger than 1 page, upload as separate excel file)
\item Captions for Movies S1 to Sx
\item Captions for Audio S1 to Sx
\end{enumerate}

\noindent\textbf{Introduction}


\noindent\textbf{Text S1.}
%


\noindent\textbf{Data Set S1.} 


\noindent\textbf{Movie S1.} 


\noindent\textbf{Audio S1.} 
\end{article}
\clearpage